\documentclass[aps,prl,reprint]{revtex4-1}
\usepackage{amsmath}
\usepackage{amssymb}
\usepackage{graphicx}
\usepackage{mathrsfs}
\usepackage{subfigure}
\newcommand{ \braces} [1] { \left( #1 \right)}
\newcommand{ \sbraces} [1] { \left\lbrace #1 \right\rbrace}

\newcommand{\nach} [2] {\frac{\partial #1}{\partial #2}}
\newcommand{\nachtot} [2] {\frac{\mathrm{d} #1}{\mathrm{d} #2}}

\newcommand{\von} [1] {\! \braces{#1}}

\begin{document}

\title{Hamiltonian active particles in an environment}

\author{Timo Eichmann}
\author{Diego Fieguth}
\author{Daniel Brady}
\author{James R. Anglin}

\affiliation{\mbox{State Research Center OPTIMAS and Fachbereich Physik,} \mbox{Technische Univerit\"at Kaiserslautern,} \mbox{D-67663 Kaiserslautern, Germany}}

\date{\today}

\begin{abstract}
We examine a Hamiltonian system which represents an active Brownian particle that can move against an external force by drawing energy from an internal depot while immersed in a noisy and dissipative environment. The Hamiltonian consists of two subsystems, one representing the active particle's motion and the other its depot of `fuel'. We show that although the active particle loses some of its energy to dissipation from the environment, dissipation can also help to stabilize the dynamical process that makes the particle active. Additionally we show how a Hamiltonian active particle can harvest energy from its environment.
\end{abstract}

\maketitle

\section{I. Introduction}
\subsection{I.1 Active Brownian particles}
Microscopic motors are currently a major subject of investigation in all natural science: biology, chemistry, and physics. Motor proteins like Kinesin, Myosin or Prestin seem to play roughly analogous roles in living cells to automobiles in industrial economies, but the microscopic versions certainly work differently. They operate out of equilibrium and thermal fluctuations are not negligible for them \cite{Broken, Noisy}. Some proposed theoretical models for microscopic motors, such as ratchets or Brownian motors, are \emph{passive} \cite{BrownMotor, Ratchet}, in the sense that they obtain their energy from fluctuations in immediately surrounding reservoirs. Other models are \emph{active}, either by being driven by systematic external forces from light or sound \cite{Extone,Exttwo}, or by drawing on an internal energy depot \cite{Design, ActStat}.

A frequently invoked model for this latter class of active microscopic motors is the \emph{active Brownian particle} \cite{Complex, Uphill, ActiveReview, ActiveMotion, Canon, DirMot, ChemFree, Assym}. In this model a particle is subject to the usual Brownian dissipation and noise, representing the surrounding medium in which the particle moves, but the particle can also propel itself through this surrounding medium, by drawing on some form of internally stored energy (the `depot'). Active Brownian particles are intended to model such real systems as motor proteins or bacteria that swim with flagella, so the energy depot of an active Brownian particle cannot be as simple as the tightly coiled spring of a wind-up toy. Precisely because the microscopic details of how microscopic motors store and use energy are complex, however, the paradigm of the active Brownian particle represents energy use and storage phenomenologically. Simple equations of motion are postulated which involve an abstract sort of internal `fuel tank' and lead to evolution that sufficiently resembles powered travel; the focus is more on what active particles can do than on just how they do it.

The use of active Brownian particles as phenomenological models actually goes well beyond molecular machines, especially when it extends to include interactions between active particles \cite{selfdriven, ActiveSus, ActiveReview, Information, Coop}. Active particles can serve as simple models for entire living organisms \cite{Active_ComplexEnvironment, Information} or even as candidates for a minimal model of cognition \cite{Cognition}. Within physics, the concepts of active particles have been helpful in developing stochastic thermodynamics \cite{Seifert, Limits, ActiveEntropy, ActStoch, Stochastic_Energetics}, and they represent the simplest limit of `active matter' in general \cite{Challenging, Need, Opinion, ActStat}.

The phenomenological `black box' treatment of \emph{how} internally stored energy is transferred into motion by an active particle has thus been a valuable feature of the active particle concept, permitting insights into a wide range of phenomena. Exactly how it is that densely stored energy can enable motion is still a basic physics question, however, and for many of the systems that are modeled with active Brownian particles the microscopic mechanisms have yet to be understood.

\subsection{I.2 Hamiltonian active particles}
When a macroscopic motor operates within a dissipative environment, we can generally take it for granted that the motor's strong and massive components will continue to move as they are designed to move, with only some power loss to friction; it is not obvious, however, that the much smaller and lighter mechanism of a microscopic engine will even continue to run under viscous drag and collisional battering. Since the simple but peculiar ``Hamiltonian daemon'' dynamical system of \cite{ClassicalDaemon} resembles a combustion engine in its behavior, yet has only a single non-trivial degree of freedom, it offers a minimalistic Hamiltonian realization of an active particle. By seeing how the ``daemon'' is affected by dissipation and noise, we can identify general issues that may be involved in the operation of more realistic Hamiltonian active particles.

Our results will offer two insights for more general active particles. We will find that dissipation from the environment can actually \emph{enhance} the efficiency of the active particle: even quite weak dissipation can allow the active particle to convert more of its internally stored energy into work, by helping to control its internal dynamical transitions. And we will confirm that even a simple Hamiltonian mechanism can harvest energy from its surroundings, acting as a sort of mindlessly automatic Maxwell's demon.

\subsection{I.3 Organization}
Our paper is structured as follows. In Section II we briefly review the Hamiltonian daemon system from Ref.~\cite{ClassicalDaemon} in the absence of any environment, to show how this dynamical system can represent an active particle. We first describe the system's engine-like behavior and explain the basic issues which it will illuminate as a toy model for an active Brownian particle. Then we present the daemon's Hamiltonian and show how to understand its behavior in terms of adiabatic orbits in a two-dimensional phase space.

In Section III we then insert our Hamiltonian active particle into an environment consisting of gas particles that collide with the active particle, inducing stochastic jumps in its momentum, but without directly affecting its energy depot. 

In Section IV we will discover and quantify our first major result, showing that although dissipation does lead to some fraction of the particle's internal energy being wasted as heat, and excessively strong dissipation can prevent the particle from being active at all, total energy transfer from internal fuel into work can actually \emph{increase} due to moderate dissipation. The environment is not simply the active particle's dissipative adversary: the environment can help to control the dynamical process that powers the active particle.

In Section V we then replace the dissipative ``buffer gas'' environment with an environment that interacts with the active particle's energy depot. We find a regime in which this kind of environment acts as a gain medium for the particle, restoring its internal energy faster than the energy depot is being drained to power the particle's motion against friction. This provides a Hamiltonian realization of the often-modeled scenario in which an active particle harvests energy from its environment. 

We conclude in Section VI with a summary and an outlook toward future work.

\section{II. A Hamiltonian active particle}
\subsection{II.1 Daemon behavior}
\begin{figure}[thb]
	\centering
	\includegraphics*[trim=59 161 23 212 , width=0.45\textwidth]{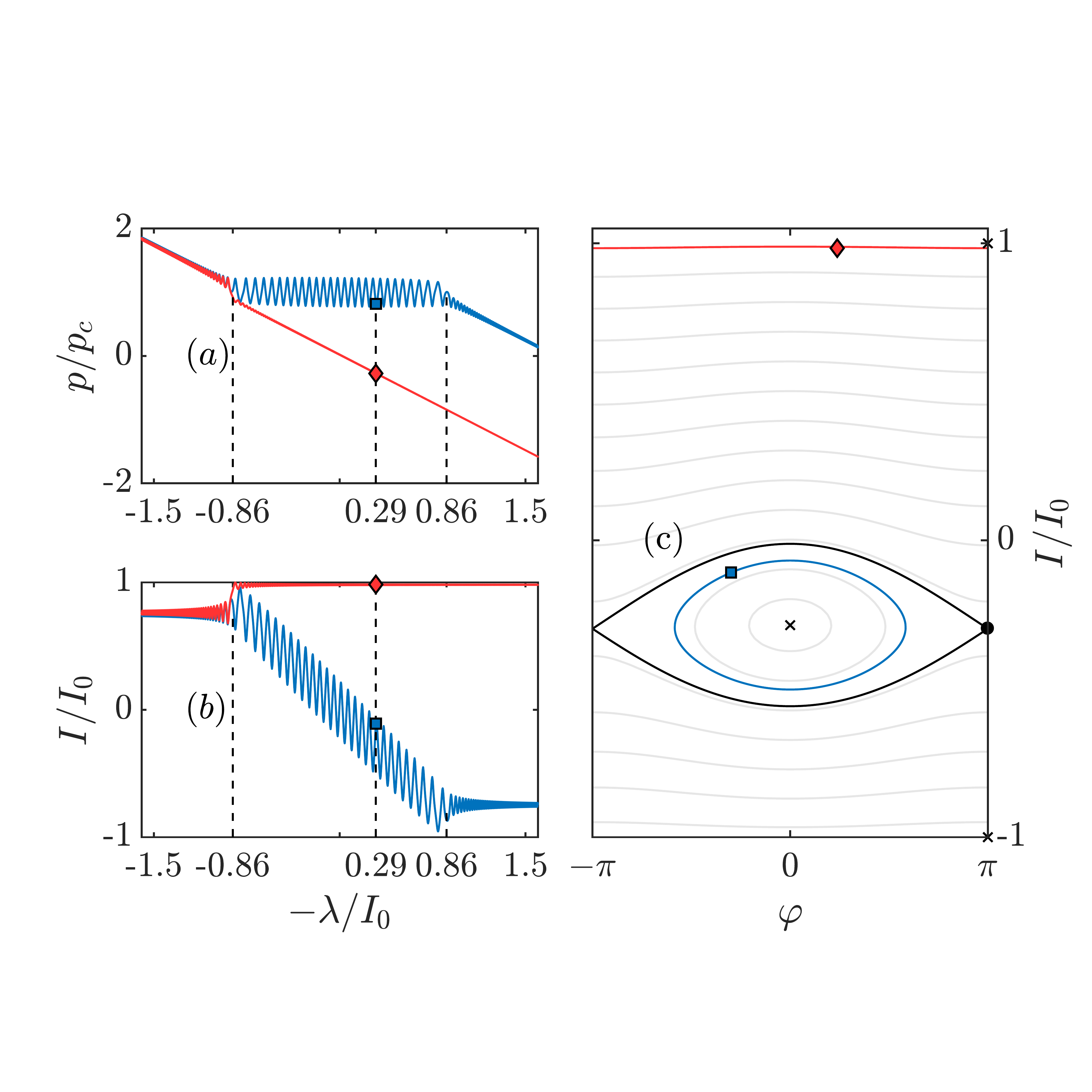}
	\caption{\label{fig:demoplot1} {\bf Left panels}: The evolution of the canonical momentums $p$ and $I$ generated by the Hamiltonian \eqref{eq:H_4dim} plotted over rescaled time $\lambda(t)$ (defined in \eqref{eq:lambda}). The blue and red trajectories have only slightly different initial values of the angle coordinate $\alpha$, but these two trajectories show the two dramatically different dynamical phases of the system. In the red case the motion is almost the same as if the fuel and motional subsystems were decoupled, whereas in the blue case most of the energy of the fuel depot is used to move the particle against the external force.
{\bf Right panel}: The phase space $\braces{\varphi,I}$ of the transformed Hamiltonian \eqref{eq:H_4dim_realization_fuel} for the rescaled time $-\lambda/I_0=0.29$. The red diamond and blue square correspond to the trajectories shown in the left panels, at the time marked there with diamonds and squares. In (c) the instantaneous energy contours on which the system is moving at that instant in the two trajectories are marked in red and blue, respectively. The closed black contour is the instantaneous separatrix that marks the division in phase space between active and inactive dynamical phases.}
\end{figure}

Before it is coupled to any environment, our Hamiltonian daemon model for an active particle behaves as shown in Fig.~\ref{fig:demoplot1}(a) and (b). Like a physical pendulum which can either swing fully around and around or else swing back and forth, or like a motor which can be running or not, our system has two dynamical phases. Trajectories which include the ``active phase'', in which the particle draws energy from its depot to move against an external force, are shown in blue; the trajectories shown in red involve only the ``inactive phase", in which the depot energy remains untapped and the particle moves ballistically. In particular Fig.~\ref{fig:demoplot1}(a) shows the particle's momentum $p$ versus the rescaled time, which decreases steadily under the external force during inactive phases, but oscillates around a constant positive momentum $p_\text{c}$, moving steadily against the external force, in the active phase. Figure~\ref{fig:demoplot1}(b) shows correspondingly that the internal energy in the depot degree of freedom remains essentially constant during inactive phases, but steadily drops to power the active phase motion.
These two dynamical phases, and the possibility of transitions between them, are the central theme of this paper. Figure~\ref{fig:demoplot1}(c) shows a two-dimensional reduced phase space, to be constructed below, in which the two kinds of orbit that represent the two phases can be recognized, divided from each other by a separatrix, like in the pendulum case former mentioned. 
Figure~\ref{fig:demoplot1} was generated using the Hamiltonian \eqref{eq:H_4dim}, which will be explained in the following, with the dimensionless parameters $M=\Omega=I_0=k=1$, $g=0.001$ and $\varepsilon=0.02$. The initial values at $-\lambda_i/I_0 = -1.75$ are $kq_i = 0$, $p_i/p_c=2$ and $I_i/I_0=0.75$. The only difference between the blue and the red curve in Fig.~\ref{fig:DemoCollision} is the initial value for $\alpha$, which is $\alpha_i=0$ ($\alpha_i=-\pi/2$) for the blue (red) curve.
As explained in \cite{ClassicalDaemon}, the isolated daemon can be understood in terms of (mostly) adiabatic evolution within this reduced phase space. This we now briefly review.

\subsection{II.2 The active particle Hamiltonian}
The daemon Hamiltonian of \cite{ClassicalDaemon} is a sum of terms representing the motion of the particle, the depot, and the coupling between them:
\begin{align} \label{eq:H_4dim}
H\von{q,p,\alpha,I} &= H_\text{M}\von{q,p} + H_\text{D}\von{\alpha,I} +\varepsilon H_\text{C}\von{q,\alpha,p,I},
\end{align}
with a small dimensionless parameter $\varepsilon$ multiplying the coupling term $H_C$, so that the coupling will transfer energy slowly from the depot into motion and thereby model an active particle rather than a rocket or bomb.
The motion and depot terms in $H$ are ordinary:
\begin{align} \label{eq:H_4dim_realization_work}
&H_\text{M} = \frac{p^2}{2M} + fq, \\ \label{eq:H_4dim_realization_fuel}
&H_\text{D} = \Omega I\;.
\end{align}
$M$ is the mass of the active particle, which can move in the one dimension of position $q$ with momentum $p$, subject to the constant external opposing force $-f$. The fuel depot $H_\text{D}$ is assumed for simplicity to be proportional to a single action variable $I$, with the dynamical frequency of the fuel degree of freedom being $\Omega$. We stipulate that $\Omega$ is by far the highest frequency in our problem, in order to represent the depot energy as densely stored (large energy for small mass and size)~\cite{ClassicalDaemon}.

\subsection{II.3 Nonlinear coupling}
The special coupling $H_\text{C}$ allows the daemon to achieve secular energy transfer from $H_\text{D}$ to $H_\text{M}$ in spite of the large frequency gap $\Omega$ and weak coupling $\epsilon$:
\begin{equation}\label{eq:H_4dim_realization_coupling}
H_\text{C} = -\Omega\sqrt{I_0^2-I^2} \cos\von{kq-\alpha}\;,
\end{equation}
where $k$ is a constant with units of inverse length, $\alpha$ is the canonically conjugate angle variable to the fuel depot action variable $I$, and the constant $I_0$ is a bound on the fuel depot action value such that $-I_0 \leq I \leq I_0$. 

This $H_\text{C}$ (\ref{eq:H_4dim_realization_coupling}) provides an effectively resonant coupling (a \emph{Chirikov resonance}~\cite{Chirikov}) between the motion of the active particle and its fuel depot, as long as the particle moves at close to the critical speed $v_\text{c}=\Omega/k$, since then $kq\sim \Omega t$ and $\cos(kq-\alpha)$ becomes slowly changing in time instead of rapidly oscillating. Unlike a typical Chirikov resonance, in which the long-term effect of the resonant couple is limited by nonlinear stabilization \cite{Chirikov,ClassicalDaemon}, this resonance can sustain itself over long times: the internally generated force on the particle from $H_\text{C}$ keeps the particle speed near $v_\text{c}$ (momentum near $p_\text{c} = M\Omega/k$) while the power to sustain this speed against the external force $f$ (and as we will see below, any viscous force) is steadily drawn from the depot energy $H_\text{D}$.

Whether the Chirikov resonance actually does sustain active motion in this way is a dynamical question within the Hamiltonian mechanics of \eqref{eq:H_4dim}. As we will now explain, the Hamiltonian active motion only persists within the central ``eye'' region, bounded by a separatrix, of the effective two-dimensional phase space shown in Fig.~\ref{fig:demoplot1}(c). In the absence of any environment the isolated system described by \eqref{eq:H_4dim} can remain within this separatrix, performing active motion against the external force, for a certain finite time which is generally \emph{shorter} than the limit imposed by energy conservation alone \cite{ClassicalDaemon}. Our question in the present paper is how environmental dissipation and noise may change the duration of active motion by expelling the system from the active region sooner---or keeping it in the active region longer.

\subsection{II.4 Effective 2D phase space} 
The Hamiltonian \eqref{eq:H_4dim} can be simplified by a canonical transformation from $(q,p),(\alpha,I)$ to the new variables $(q,J),(\varphi,I)$ using
\begin{align} \label{eq:J}
%\varphi &= \alpha - kq, & J =p+kI+gt,
\alpha\to\  &\alpha(\varphi,q)=\varphi + kq,\nonumber\\
p\to\  &p(J,I,t)= J-kI-ft
\;.
\end{align}
This simplifies the system because $J=p +kI+ft$ is an exact constant of the motion (even though it  depends explicitly on time $t$!): $\nachtot{J}{t}=\sbraces{J,H}+\nach{J}{t}=0$. 
Taking the time-dependence of the canonical transformation properly into account, the Hamiltonian in the new coordinates is 
\begin{align}
{H'} &= \frac{[p(J,I,t)]^2}{2M}+\Omega I  - \varepsilon \Omega \sqrt{I_0^2-I^2} \cos\von{\varphi} \notag\\
&\equiv \frac{k^2}{2M}\Bigl[I-\lambda(t)\Bigr]^2 - \varepsilon \Omega \sqrt{I_0^2-I^2} \cos\von{\varphi} \notag\\
&\quad + E_0(J,t), \label{eq:H_2dim}
\end{align}
where
\begin{align}
E_0(J,t) &= -\frac{M\Omega^2}{2k^2}-\frac{f\Omega t}{k}+\frac{\Omega}{k}J\nonumber,\\
\lambda(t) &= \frac{f}{k} (t_0-t),\nonumber\\
t_0(J)&=\frac{Jk-M\Omega}{fk}\label{eq:lambda}\;.
\end{align}

The evolution of $(q,J)$ is thus trivial once $\varphi(t)$ and $I(t)$ have been found from equations of motion that do not involve $q$. For these the term $E_0(J,t)$ is irrelevant and can be ignored, while the effective shift in the time origin $t_0(J)$ is likewise trivial. Our system without dissipation is thus exactly represented as a time-dependent rotor model, even though the full Hamiltonian, with its two degrees of freedom, is time-independent. 

\subsection{II.5 Adiabatic regime}
The maximum force which $H_\text{C}$ can exert on the particle is $\varepsilon k \Omega I_0$, and so the particle can only make its way actively against the opposing force if $f < \varepsilon k \Omega I_0$; we assume this from now on. We also note that getting the active particle up to the critical speed $v_c$ requires an initial energy input $Mv_\text{c}^2/2$, and so it will only be worthwhile trying to exploit our system's energy depot if $\Delta \leq 1$, with
\begin{align} \label{eq:Delta}
\Delta^2 = \frac{M\Omega^2/k^2}{\Omega I_0}.
\end{align}

The typical rate at which the Hamiltonian $H'$ changes is $f/(\Omega k I_0)$, while the frequency of orbits around the instantaneous fixed point at $\varphi=0$ and $I = - f(t-t_0)/(\Omega k)$ is $(\varepsilon \Omega I_0 k^2/M)^{1/2}$. The ratio of these two rates is therefore small in all regimes we consider, and so the adiabatic theorem of classical mechanics tells us that our effective rotor system closely follows orbits of constant instantaneous $H'=E$, even as these slowly move and deform while $H'$ changes. Such instantaneous orbits are shown in Fig.~\ref{fig:demoplot1}(c). The slow (adiabatic) motion and deformation of the orbits depends on time only through the parameter $\lambda(t)$ defined in \eqref{eq:lambda}; the dimensionless ratio $-\lambda(t)/I_0$ can be considered as a conveniently rescaled time variable.

For our particular $H'(\varphi, I, t)$, the eye-shaped separatrix migrates slowly but steadily downward in $I$, so that on all orbits inside the separatrix the depot steadily loses energy, sustaining active motion at a nearly steady speed of $v_\text{c}$. Outside the separatrix, in contrast, orbits merely flatten or bend slightly over time, leaving $I$ nearly constant and providing no power to move the particle actively.

As well as moving slowly downward in $I$, the separatrix also slowly changes in shape and size. In particular, as the mean $I$ inside the separatrix approaches the depot ground state $I=-I_0$, the $\sqrt{I_0^2-I^2}$ factor makes the area enclosed by the separatrix shrink to zero. This is decisive, because under adiabatic conditions the phase space area enclosed by the system's actual orbit stays constant (\emph{adiabatic invariance}). The system can therefore only remain inside the separatrix, keeping the particle active, as long as the separatrix area is larger than its conserved orbit area. Once the separatrix shrinks smaller than this, the system exits the separatrix and the particle goes inactive, even if $H_\text{D}$ still holds enough energy to power more active motion. 

We refer readers to \cite{ClassicalDaemon} and \cite{Liouvillepaper} for further details on these mechanisms. The question for the present investigation is how the addition of environmental dissipation and noise, which modify both Liouville's theorem and the adiabatic theorem, can alter the behavior of this Hamiltonian active particle system.
  
\section{III. Hamiltonian active particle in an environment: Stochastic collisions}

We now consider that our active particle moves through a (one-dimensional) dilute gas of light particles having mass $m \ll M$. Elastic collisions occur at random times as the continuous limit of a Poissonian sequence, \textit{i.e.} with an exponentially decaying probability distribution of durations $\tau$ between collisions, $P(\tau) = \bar{\tau}^{-1}e^{-\tau/\bar\tau}$ for the average collision rate $1/\bar{\tau}$. Collisions will alter the active particle's momentum instantaneously; between collisions the system evolves under \eqref{eq:H_4dim} as in Section II above. The full evolution of the system is thus a piecewise deterministic process \cite{Piecewise_deterministic} which can easily be simulated, and so we will begin our investigation of Hamiltonian active Brownian particles by exploring the qualitative effects of a Brownian environment on our Hamiltonian active particles within this collisional model. After identifying a potentially dramatic and counter-intuitive effect in which environmental dissipation can actually help the active particle travel farther than it otherwise would, we will simplify our model to pure viscous damping in order to obtain some more general results.

\subsection{III.1 Including collisions in the model}
An elastic collision between the active particle with mass $M$ and an environment particle with mass $m$ changes $p$ to $p+\delta p$ with
 \begin{equation}\delta p =\mu \braces{Mv-p},\label{eq:Delta p}
 \end{equation} 
where $\mu=\frac{2m}{M+m}\ll 1$. The velocity $v$ of an environment particle is a Gaussian random variable with a standard deviation $\sigma$, representing a Maxwell-Boltzmann distribution in 1D. 

\begin{figure}[htb]
	\centering
	\includegraphics*[trim=59 162 23 212, width=0.45\textwidth]{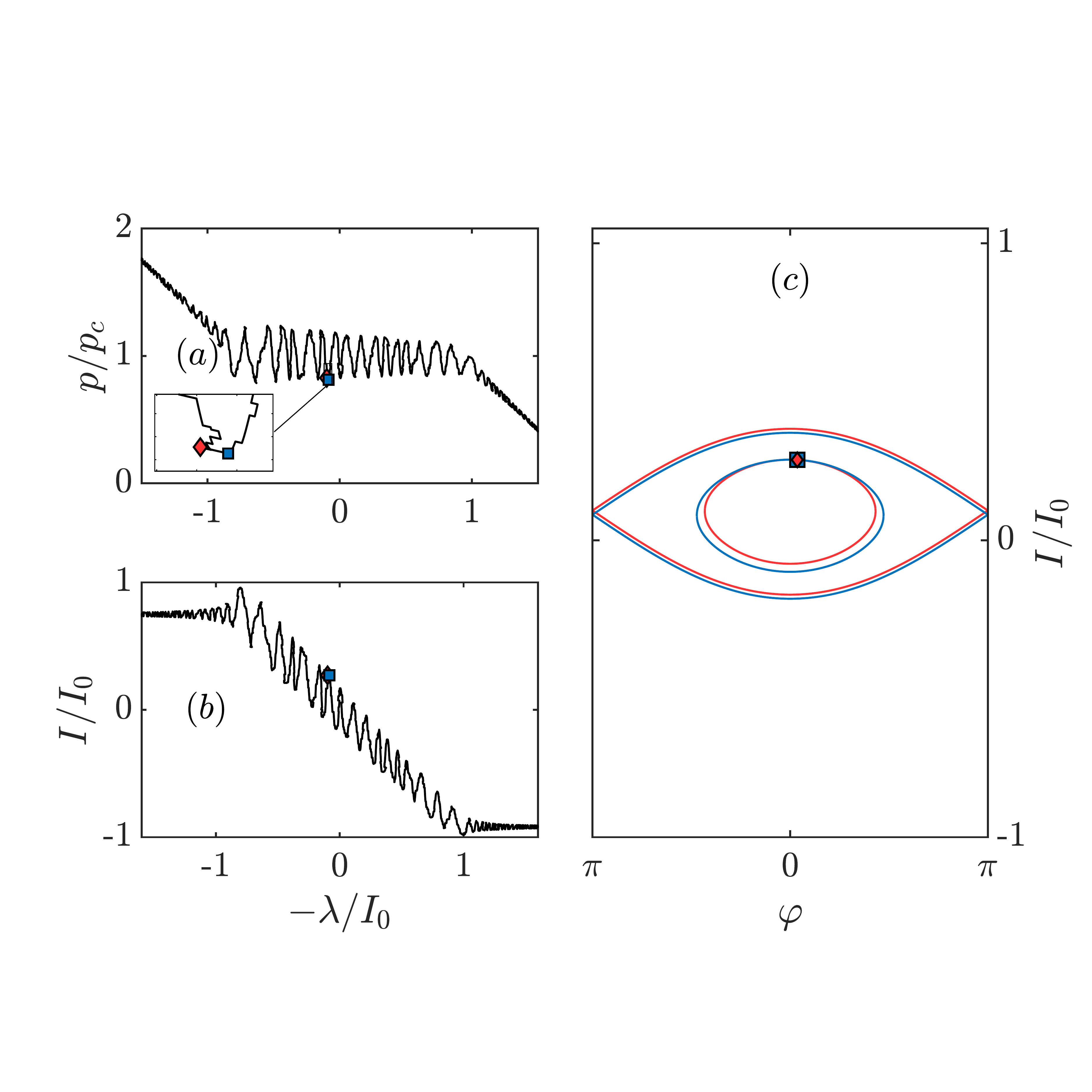}
	\caption{\label{fig:DemoCollision} Collisions lead to random jumps in the momentum $p$ and as a consequence to the random behavior of $I$. The coupling $H_\text{C}$ manages nevertheless to keep the momentum near the critical value $p_c$ during the running phase, as can be seen in (a), while $I$ decreases on average linearly (b). In the reduced phase space a collision leads to a shift of the separatrix as can be seen in (c). In (a) the red (blue) dot corresponds to before (after) a collision. In the reduced phase space shown in (c) both dots share the same location but belong to different energy contours. The horizontal axis variable $-\lambda(t)/I_0$ is now only a monotonic function of $t$ on average: the inset in (a) shows the small horizontal jitter in the parametric curve $(\lambda(t),p(t))$, as explained in the text.}
\end{figure}

To see what an environment can do to our Hamiltonian active particle, therefore, we randomly generate a sequence of collision times at average rate $1/\bar\tau$, at each of which momentum jumps with random $v$ are applied. Between collisions Hamiltonian evolution under \eqref{eq:H_4dim}-\eqref{eq:H_4dim_realization_coupling} rules instead. A typical evolution is shown in Fig.~\ref{fig:DemoCollision}, which can be compared directly with Fig.~\ref{fig:demoplot1}---though the comparison is in some ways not trivial. Figure~\ref{fig:DemoCollision} was generated with the same parameters as Fig.~\ref{fig:demoplot1} and the additional collision parameters $\sigma=2\frac{p_\text{c}}{M}$, $\mu=\frac{2}{1001}$ and $\overline{\tau}=\mu \frac{kI_0}{f}$.
A somewhat unusual feature of the horizontal axis variable $-\lambda/I_0$ in Fig.~\ref{fig:DemoCollision} is that it is no longer a simple linear function of time $t$, as it was in Fig.~\ref{fig:demoplot1}. The definition is still 
\begin{align}
-\frac{\lambda(t)}{I_0} &= \frac{f}{ k I_0} t + \frac{M\Omega}{I_0 k^2}-\frac{J}{k I_0},\nonumber\\
J& = p +kI + ft\;,
\end{align}
and $J$ is still exactly constant between collisions. Each instantaneous collision when it occurs, however, gives a (partly) random kick to $p$; since $I$ is not directly changed in a collision, the effect on the evolution of $\varphi$ and $I$ is to give $J$ a random shift in each collision. The dimensionless ratio $-\lambda(t)/I_0$ therefore increases in constant linear proportion to $t$ between collisions, but makes small a random jump (which can even be backward!) at each randomly timed collision. These effective small jitters in ``time'' (\emph{i.e.} in $-\lambda(t)/I_0$) can be seen in the inset of  Fig.~\ref{fig:DemoCollision}(a). This quasi-Brownian motion in $\lambda(t)$ provides a Brownian component in $\dot{\varphi}$ under $H'$, so that our active particle is indeed Brownian, even in the reduced phase $(\varphi,I)$ phase space.

On the larger scale that is easily visible in Fig.~\ref{fig:DemoCollision}, nonetheless, the overall behavior of the Brownian active particle remains quite similar to the case without collisions with gas particles. An active phase of the dynamics still exists, in which the particle moves for a certain time with a nearly constant momentum $p_c$, doing work against the external force while the energy in the internal depot decreases. The self-sustaining Chirikov resonance that is the minimalistically simple mechanism for this Hamiltonian active particle may be an exotic piece of nonlinear dynamics but it is not especially fragile. It can still power an active particle with Brownian noise and dissipation turned on.

\subsection{III.2 Viscous drag}
One significant effect of the dissipative environment on the active particle is not apparent in Fig.~\ref{fig:DemoCollision}: some of the power being drawn from the depot must now be used to overcome viscous damping, as the active particle moves through the gas. This basic effect does not show up in the decrease of $\Omega I$ with $-\lambda/I_0$, however, but rather in making $-\lambda/I_0$ increase faster, on average, with $t$. 

This effect can be seen from the fact that between collisions we have $-\dot\lambda(t)/I_0 = f/(k I_0)$, while according to \eqref{eq:Delta p} the average change of $-\lambda(t)/I_0$ in a collision is $-\mu p_\text{c} /(k I_0)$ in the active phase (since then $p$ stays close to $p_c = M\Omega/k$). With collisions occurring at the average rate $1/\bar{\tau}$, the average increase rate in $t$ of our rescaled ``dimensionless time'' is therefore
\begin{align}
\left\langle\frac{d}{dt}\left(-\frac{\lambda(t)}{I_0}\right)\right\rangle &= \frac{f+\gamma p_\text{c}}{k I_0},
\end{align}
where
\begin{align}
\gamma &= \frac{\mu}{\bar{\tau}}
\end{align}
can be recognized as the damping coefficient for our one-dimensional ideal gas dissipation model, inasmuch as for $\varepsilon\to 0$ we would have the average time dependence $\dot{p}=-f-\gamma p$. 

\begin{figure}[htb]
	\centering
	\includegraphics*[trim=211 317 282 357, width=0.45\textwidth]{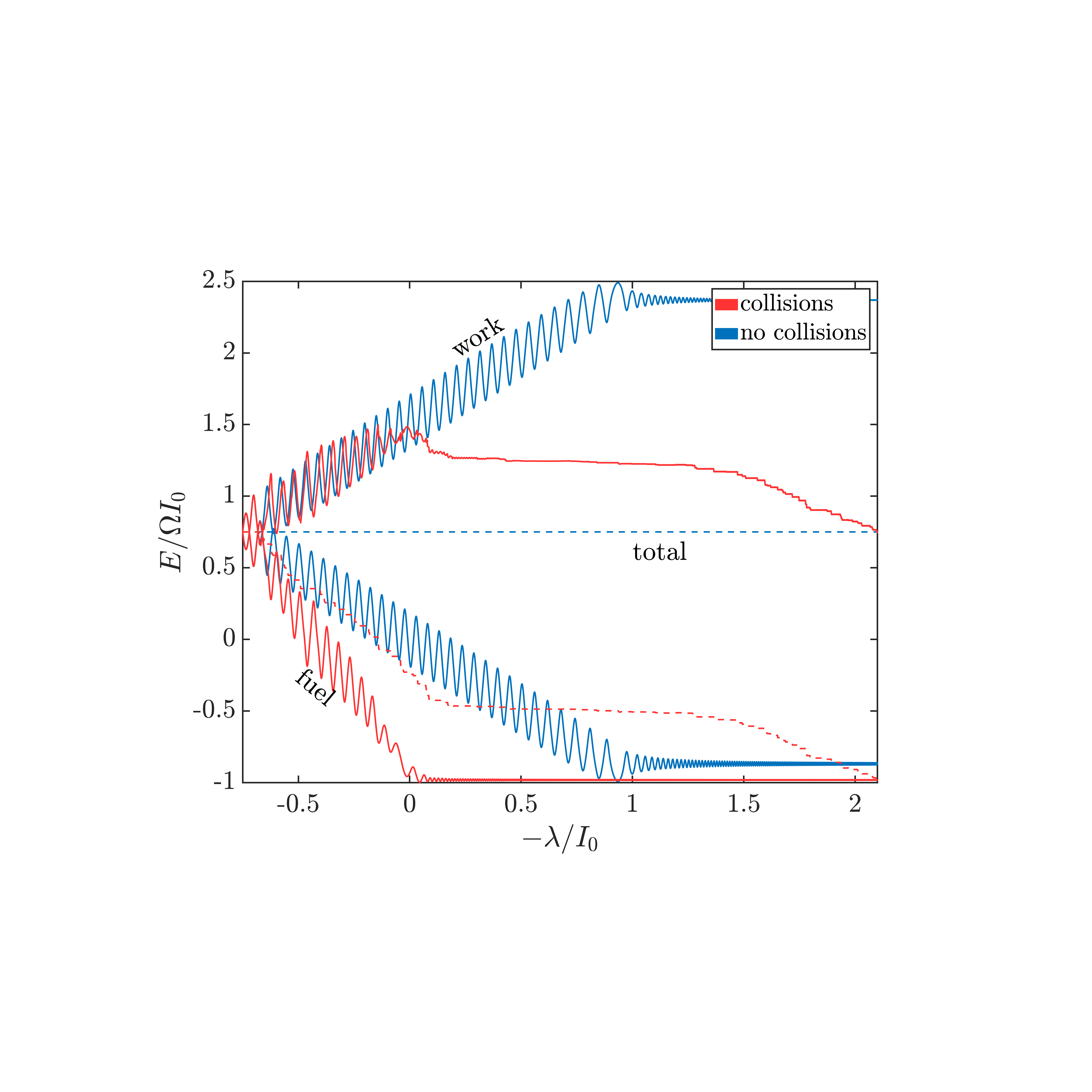}
	\caption{\label{fig:energies} Comparison of the energy between the cases with and without collisions, where $-\lambda/I_0$ corresponds to the case without collisions. The energy in the work term increases at the beginning similar in both cases. The energy transfer to the work term stops earlier for the case with collisions. For this case energy gets drained faster from the fuel. The total energy does not stay constant, since heat is created in the environment.
(The work and total energy have been shifted, by the same constant in all cases, for visual clarity. The work and total energy with dissipation remain rather flat for a while after $\lambda =0$, but apart from the visible Brownian fluctuations these curves do closely follow the motion of a particle decelerating from $v_\text{c}$ and finally attaining terminal velocity $-f/(\gamma M)$ in the negative direction, under constant force $-f$ and damping~$\gamma$.)}
\end{figure}

This basic effect of viscous drag can be seen directly by plotting $H_\text{M}$, $H_\text{D}$, and their sum; see Fig.~\ref{fig:energies}. In this Figure the horizontal axis is $-\lambda/I_0$ for the case without dissipation, plotted in blue for comparison; the case without dissipation is simply plotted against the linearly rescaled time $t$ that coincides with the dissipationless $-\lambda/I_0$.  The depot energy is thus depleted sooner because of gas particle collisions, although the active particle still moves against the same external force $f$ at the same average $v_\text{c}$ that is set by the Chirikov resonance. Since the collisions are elastic the missing energy has clearly gone into heating the gas. The collision parameters for the red curve are $\sigma=0.6 \frac{p_c}{M}$, $\mu=\frac{2}{101}$ and $\bar{\tau}=\mu \frac{k I_0}{f}$. Both curves start with initially $p_i/p_\text{c} = 1$. All other parameters are chosen as in Fig.~2.

\subsection{III.3 Thermal fluctuations}
\begin{figure}[htb]
	\centering
	\includegraphics*[trim=25 178 100 211, width=0.45\textwidth]{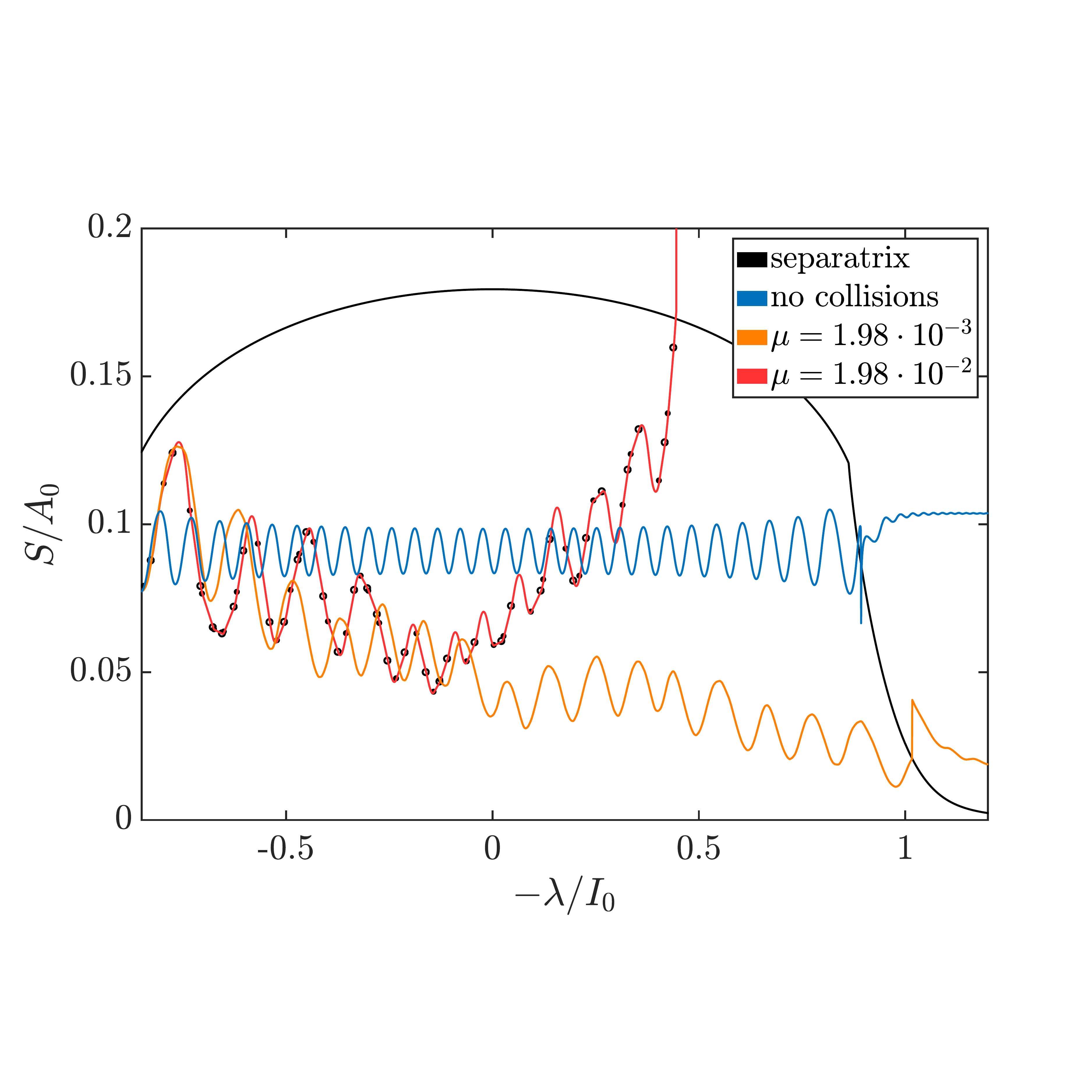}
	\caption{\label{fig:demoAction} Action for different collision parameters and the separatrix plotted over the corresponding $-\lambda/I_0$. The system changes its state from running to stalling, if the action of the orbit equals the area of the separatrix. Friction leads to a decreasing action, and with that to a lower value of $\lambda$ (orange) compared to the collision less case (blue). Strong fluctuations might lead to unpredictable transitions (red). On the red curve collisions are marked with dots.}
\end{figure}

With sufficiently mild Brownian kicking the working efficiency of the active particle is reduced by viscous drag, but the basic operation of the active particle persists. Stronger thermal fluctuations from gas collisions can indeed disrupt the Hamiltonian mechanism, however. To show this we present two different numerical evolutions in Fig.~\ref{fig:demoAction}, with the gas particle mass $m$ differing by an order of magnitude; the horizontal axis is again the dimensionless $-\lambda/I_0$ rescaled time of the dissipationless case, which is also shown again for comparison. The vertical axis here is neither $I$ nor $p$ but rather the action variable $S$, defined as the area enclosed in the $\varphi,I$ phase-space plane by the contour on which the instantaneous function $H'$ equals the instantaneous value of $H'$. The quantity $A_0=4\pi I_0$ is the total area of the phase space. The fact that in Fig.~\ref{fig:demoAction} the blue curve of the dissipationless case oscillates around a constant value represents the well-known adiabatic invariance of the action~\cite{Goldstein}. All lines correspond to the same initial values $I_i/I_0=0.85$, $p_i/p_\text{c}=1$, and $\alpha_i=\frac{5}{8}\pi$. The collision properties for the red line are $\sigma=0.15 \frac{p_c}{M}$, $\mu=\frac{2}{101}$, $\bar{\tau}=\mu \frac{k I_0}{f}$, and $\sigma=0.1 \frac{p_c}{M}$, $\mu=\frac{2}{1001}$, $\bar{\tau}=\mu \frac{k I_0}{f}$ for the orange line.
The black border in Fig.~\ref{fig:demoAction} shows the phase space area enclosed by the instantaneous separatrix within which the active phase occurs. When the system's trajectory crosses this border, the active phase ceases. We can see that for large enough gas particle mass $m$ the larger Brownian fluctuations in $p$ can include an unlucky run that drives the system outside the separatrix well before it would have exited without dissipation. For smaller $m$, in contrast, such large stochastic sorties become extremely unlikely. 

The range of possible behaviors of complex Hamiltonian mechanisms like this one under stochastic forces is too large to explore fully within the scope of this paper; here we simply note that a low-temperature regime does exist in which the stochastic effects remain small. Further studies of fluctuations, perhaps using Fokker-Planck equations, will be left for the future; we focus now instead upon a significant effect of pure dissipation apart from the simple drag and heating discussed above.

\section{IV. Dissipative control}
Even though the plots of $p$ and $I$ versus $-\lambda/I_0$ in Fig.~\ref{fig:DemoCollision} do not show the additional energy drain of the viscous drag, they do show a noticeable trend of steadily \emph{decreasing amplitude} in the \emph{oscillations} of $p$ and $I$ around the constant or linear decrease, respectively, of the active phase. A similar decrease in oscillation amplitude can also be seen in Fig.~\ref{fig:energies}; the steadily decreasing action variable in Fig.~\ref{fig:demoAction} is the same effect, since the area enclosed by orbits in phase space is proportional to the oscillation amplitude squared. The amplitude decrease in these particular cases is modest, but its potential significance can be seen in the fact that the orange curve in Fig.~\ref{fig:demoAction}, with dissipation and weak noise, exits the separatrix slightly \emph{later} than the blue curve with no dissipation. This means that the active phase actually lasts a bit longer in the case with dissipation than it would have without dissipation.

\subsection{IV. 1 Shrinking action}
With different parameter values this dissipative effect can become more dramatic. To investigate the effect systematically we replace our random collision model with a constant viscous damping, eliminating the random momentum kicks; this is equivalent to taking the limits $\mu\to 0$ and $\bar{\tau}\to0$ in our collision model while keeping $\gamma = \mu/\bar{\tau}$ fixed. We therefore return from our 2D reduced phase space to the full canonical Hamiltonian equations of motions derived from \eqref{eq:H_4dim}, with the addition of damping on the particle momentum $p$:
\begin{align}\label{dissEoM}
\dot{p} &= -f-\varepsilon k\Omega\sqrt{I_0^2-I^2}\sin(kq-\alpha)-\gamma p\;,\nonumber\\
\dot{q} &= \frac{p}{M}\;,\nonumber\\
\dot{I} &= \varepsilon\Omega\sqrt{I_0^2-I^2}\sin(kq-\alpha)\;,\nonumber\\
\dot{\alpha} &= \Omega + \varepsilon\frac{\Omega I}{\sqrt{I_0^2-I^2}}\cos(kq-\alpha)\;.
\end{align}
In the active phase of evolution we will have $p\sim M\Omega/k$ as explained above, and so active motion will only be possible if the total retarding force $f+\gamma M\Omega/k$ is less than the maximum force $\varepsilon k\Omega I_0$ that $H_\text{C}$ can exert.

In Appendix A we use multiple time scale analysis to show that for small $\sqrt{\varepsilon}$ the adiabatic invariance of the system's action variable $S$ is replaced by exponential decay at rate $\gamma$ during the running phase:
\begin{equation}
S(t) = S_i e^{-\gamma t} [1+ \mathcal{O}(\sqrt{\varepsilon})]\;,
\end{equation}
where the order $\sqrt{\varepsilon}$ corrections remain small even over long times. This surprisingly simple but non-trivial result holds even when the adiabatic orbits are of finite size and the motion in phase space is quite anharmonic.

This exponential decay of the orbit area in the $(\varphi,I)$ phase space (or equivalently of the area in the alternative 2D phase space $(q-\alpha/k,p)$) is an effect of dissipation besides the increased rate of consumption of $H_\text{D}$ due to viscous drag with coefficient $\gamma$ on the particle moving at steady speed $v_c$. The decay of $S(t)$ represents damped motion of the internal mechanism of the active particle, rather than of the motion of the particle itself; it is analogous to damping of engine vibration in a car, rather than to aerodynamic drag on the car's body. This internal damping occurs as an indirect effect of the damping that acts directly only on the active particle's linear momentum, simply because the particle's internal mechanism is coupled to its momentum---as it must be, of course, in order to power the particle's motion actively.

\subsection{IV.2 Effectiveness of internal damping}
The internal damping can potentially have a more significant effect on the system's long-term evolution than the viscous drag. The decrease in efficiency due to drag is simply the proportion of drag force in the total force opposing the active motion, $\gamma p_c /(f+\gamma p_c)$. The exponential decay of $S(t)$ continues as long as the active phase lasts, however. Over the total running time of order
\begin{equation}
t_\text{max} = \frac{I_0\Omega}{(f+\gamma p_\text{c})v_\text{c}} = \frac{I_0 k}{f+\gamma p_\text{c}}
\end{equation}
the e-fold reduction of $S$ will be of order
\begin{equation}
\ln\frac{S_i}{S(t_\text{max})} = \gamma t_\text{max} = \frac{\gamma p_\text{c}}{f+\gamma p_\text{c}}\frac{I_0 k^2}{M\Omega} = \frac{\gamma p_\text{c}}{f+\gamma p_\text{c}}\frac{1}{\Delta}\;,
\end{equation}
with $\Delta$ defined in \eqref{eq:Delta}.

In our numerical evolutions so far we have used $\Delta=1$, in order to avoid the large ratio of time scales that makes it hard to see all significant features of time evolution in a single plot. An active particle which can remain active for a long time because its depot energy is large must have large $1/\Delta$, however. The longer an active particle remains active, the more significant the exponentially compounding effects of internal dissipation become, in comparison to the simple flat rate of efficiency reduction due to viscous drag.

Such significant internal damping can actually be advantageous, because the mechanism by which a Hamiltonian active particle converts depot energy into motional work is not simply a phenomenological \emph{fiat} constrained only by energy conservation, but a deterministic dynamical process that stops and starts under its own equations of motion. The starting and stopping of the active phase for our particular Hamiltonian active particle model involves nonlinear dynamics with multiple time scales and a Chirikov resonance, but any actual active particle will have to control itself, mechanically, somehow. Dissipation can help with control.

\subsection{IV.3 Duration of the active phase}
In our particular case the starting and stopping of the active phase is determined by the adiabatic growing and shrinking of the instantaneous separatrix in phase space \cite{ClassicalDaemon,Liouvillepaper}. As shown in Fig.~\ref{fig:demoAction}, as $-\lambda/I_0$ increases the separatrix first grows and then shrinks. We consider that the active phase must start from the inactive phase; we do not consider initial conditions that are fine-tuned, by external forces not represented in our Hamiltonian, to ensure that the system starts in the active phase. Exactly how the system enters the separatrix is a non-trivial problem in post-adiabatic dynamics, because the adiabatic theorem breaks down at the separatrix \cite{ClassicalDaemon,Liouvillepaper}, but whenever the separatrix is growing there must be some orbits which enter it from outside, because Liouville's theorem tells us that the density of orbits inside the separatrix cannot decrease, and dissipation will only draw in more orbits, because the active phase is an energy basin.

Our system will therefore always begin its active phase on an orbit just inside the separatrix, for some initial value of depot fuel level $I/I_0$. The initial value of the system's action $S_i$ is therefore the area enclosed by the separatrix for $I=I_i$. The integral which defines the separatrix area $S_\text{sep}(I/I_0)$ is given in Appendix A; in the limit $f+\gamma p_\text{c}\to 0$ it approaches
\begin{align}
S_\text{sep}\von{\frac{I}{I_0}} &\doteq 16\sqrt{\frac{\varepsilon}{\Delta}}\frac{M\Omega}{k^2}\braces{1-\braces{\frac{I}{I_0}}^2}^{1/4}\;.
\end{align}
For general $f$ and $\gamma$, $S_\text{sep}(I/I_0)$ remains a well-defined function which must simply be computed by numerical evaluation of the integral given in the Appendix. It retains in general the properties that $S_\text{sep}(\pm 1) = 0$ and $S_\text{sep}(-I/I_0) = S_\text{sep}(I/I_0)$.

If the separatrix is expanding the system will remain in the active phase, on an orbit deeper inside the expanded separatrix, but as fuel is drained the separatrix must eventually shrink, disappearing completely when $I/I_0 =-1$. The system will exit the separatrix even before the depot is thus drained to its ground state, however, if the system's orbit action does not decrease fast enough to remain inside the shrinking separatrix. The depot energy level $I_f$ at the end of the active phase is thus fixed by the condition
\begin{equation}
S_\text{sep}\braces{\frac{I_f}{I_0}} = S_\text{sep}\braces{\frac{I_i}{I_0}}e^{-\gamma(t_f-t_i)}\;.
\end{equation}
If we then consider that the steady fuel drain in the active phase implies
\begin{equation}
t_f-t_i= \frac{(I_i-I_f)\Omega}{(f+\gamma p_\text{c})v_\text{c}} = \frac{(I_i-I_f)k}{(f+\gamma p_\text{c})}\;,
\end{equation}
we can conclude that the duration of the active phase is fixed by
\begin{equation}\label{condit}
S_\text{sep}\Big(\frac{I_i}{I_0}\Big) e^{-\frac{\gamma p_c}{f+\gamma p_c}\frac{1}{\Delta}\frac{I_i}{I_0}} = S_\text{sep}\Big(\frac{I_f}{I_0}\Big)e^{-\frac{\gamma p_c}{f+\gamma p_c}\frac{1}{\Delta}\frac{I_f}{I_0}}\;.
\end{equation}
The systematic multiple time scale analysis in the Appendix confirms this result, to leading order in the small parameter $\sqrt{\varepsilon}$.

The general properties that $S_{sep}(I/I_0) =S_{sep}(-I/I_0)$ and that $S_{sep}(\pm 1) = 0$ imply that for $I_i = I_0$ we always have $I_f=-I_0$, so that all depot energy is successfully consumed regardless of $\gamma p_c/f$, while for $\gamma=0$ and $I_i<I_0$ we have $I_f = -I_i$ (and the active phase can never begin for $I_i < 0$ \cite{ClassicalDaemon}). For $I_i < 1$ the active particle without dissipation is inefficient, in that it is unable to use all its potentially available fuel. Whenever dissipation is strong enough that
\begin{equation}
\exp\left[-\frac{\gamma p_c}{\gamma p_c +f}\frac{1}{\Delta}\right]\doteq 0\;,
\end{equation}
however, we will have $I_f \doteq -I_0$ regardless of $I_i$, meaning that all fuel is used. When $\Delta$ is small even a very slight amount of dissipation can dramatically boost the efficiency with which the Hamiltonian active particle can exploit its energy depot, by extending the duration of the active phase.

\subsection{IV.4 Efficiency}
\begin{figure}[htb]
	\centering
	\includegraphics*[trim=232 316 261 357, width=0.45\textwidth]{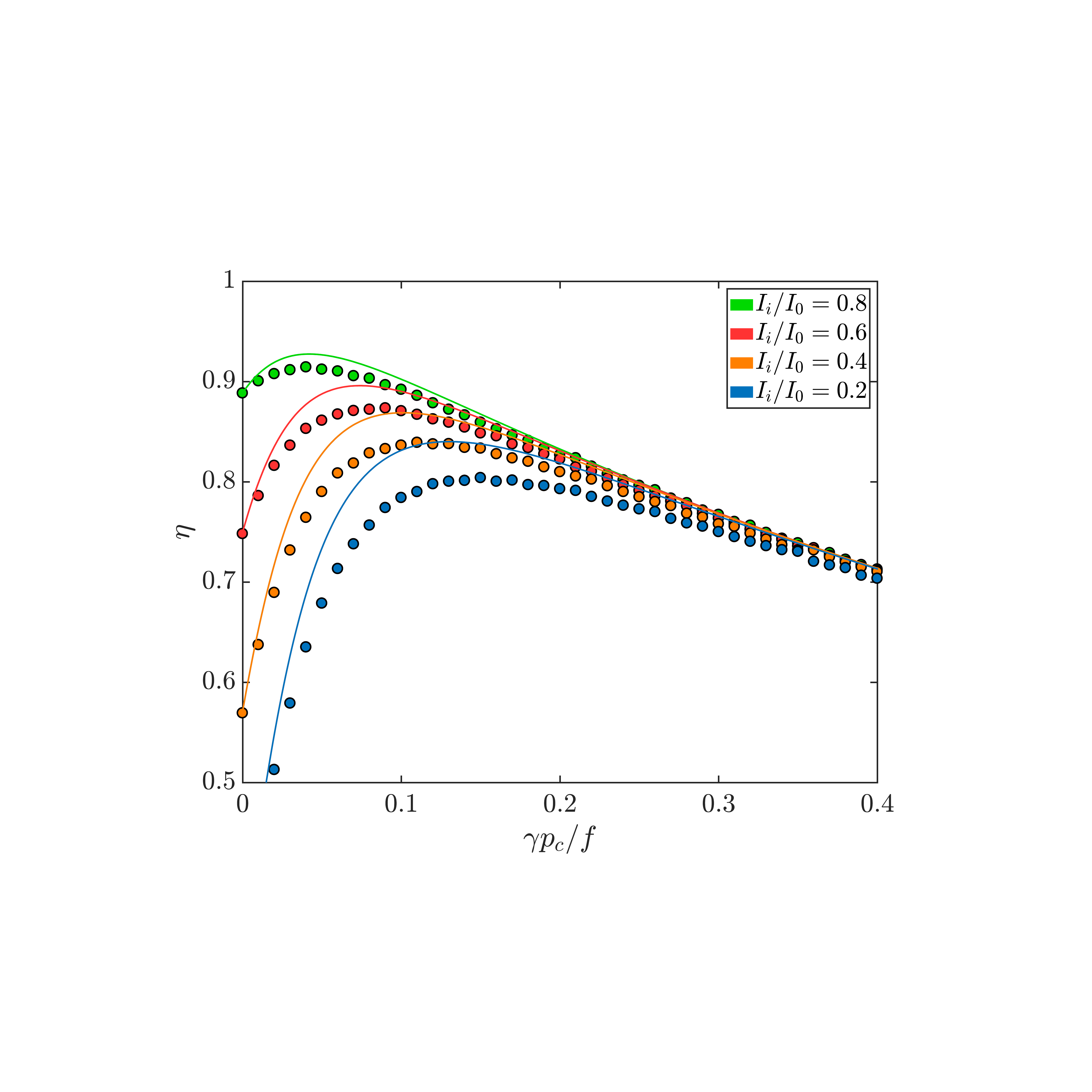}
	\caption{\label{effic1} Efficiency $\eta$ versus $\gamma p_c /f$ for different initial values $I_i$. Dots are from numerical evolutions under the equations of motion; the curves are the analytical predictions as explained in the text. Efficiency can be boosted by dissipation significantly.}
\end{figure}

To measure this effect we define the total efficiency $\eta$ of the active particle as the fraction of depot energy which is successfully consumed, times the fraction of power which goes into motional work against the external force $f$ rather than being expended against drag and turned into heat:
\begin{equation}
\eta = \frac{I_i-I_f}{I_i + I_0}\frac{f}{\gamma p_c +f}\;.
\end{equation}
The analytical predictions for this efficiency are plotted for different $I_i$ in Fig.~\ref{effic1}, as a function of $\gamma p_c /f$, for a case in which the multiple time scale theory should be very good because $\varepsilon = 10^{-4}$, and in which the efficiency boost from dissipation can be substantial because $\Delta = 1/4$. Also shown is the same efficiency for a single numerical solution of the equations of motion \eqref{dissEoM}, from an initial condition just inside the initial separatrix. The numerically exact evolution generally falls a few percent short of the analytically predicted efficiency; this is what can be expected for $\sqrt{\varepsilon}=0.01$.

\begin{figure}[htb]
	\centering
	\includegraphics*[trim=233 316 263 357, width=0.45\textwidth]{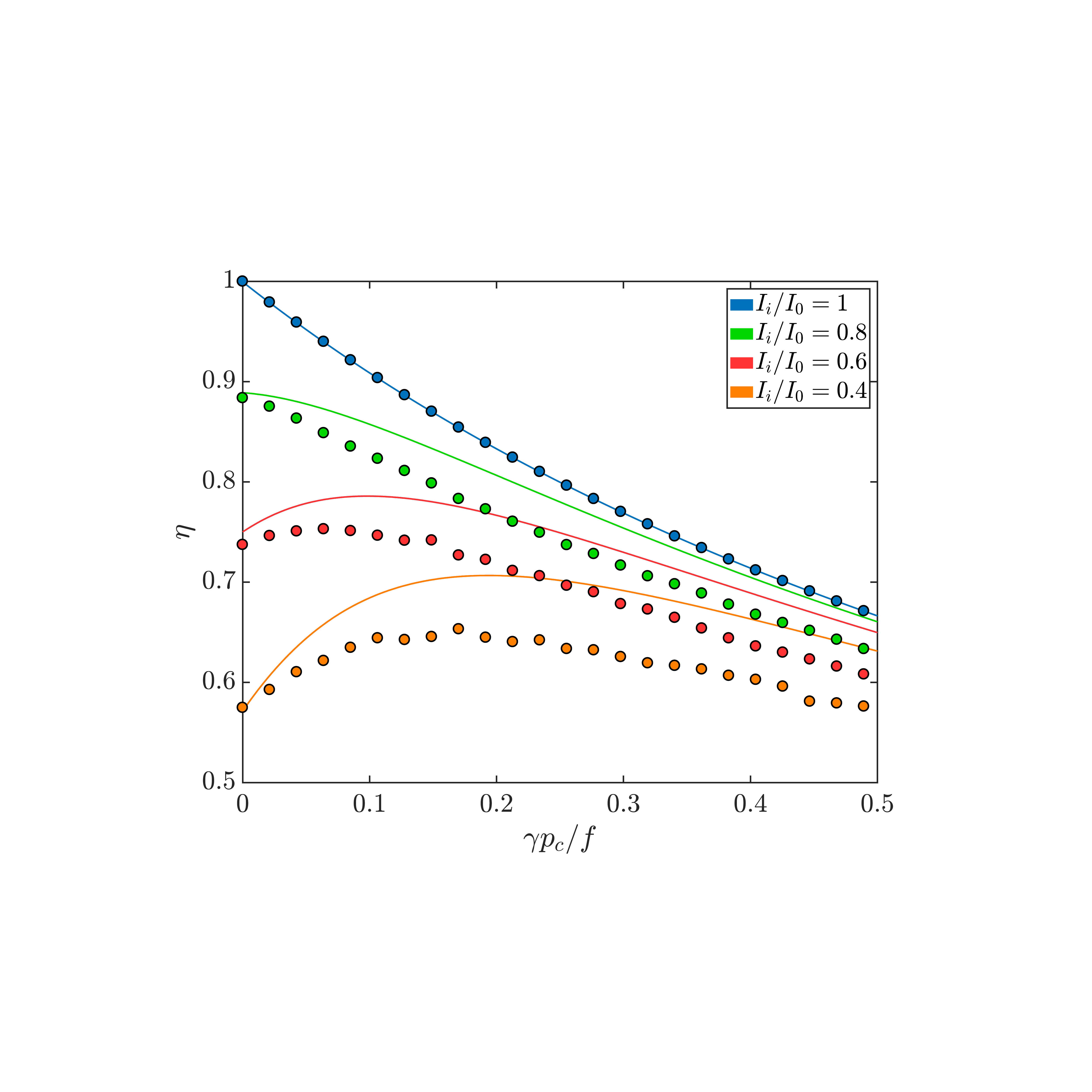}
	\caption{\label{fig:demoEfficiency} Efficiency $\eta$ versus $\gamma p_c /f$ for different initial values $I_i$ with less adiabatic parameters than in Fig.~\ref{effic1}.}
\end{figure}

A less extreme case with the same parameters $\varepsilon = 0.2$ and $\Delta = 1$, as in the plots of our previous Sections, shows that dissipation can still increase efficiency even when conditions are not so ideal, see Fig.~\ref{fig:demoEfficiency}. At this considerably larger value of $\sqrt{\varepsilon}$ the post-adiabatic corrections are larger, and the dependence of efficiency on precise initial conditions (just where the system enters the separatrix at the start of the active phase) is significant. The points in Fig.~\ref{fig:demoEfficiency} are therefore the average efficiencies of 100 different numerical evolutions with random initial conditions for $\alpha$. 

Even in this much less extreme case the dissipatively enhanced control of the dynamical process which powers the active particle can more than compensate for the losses to drag. This enhancement will not matter if the fuel depot can reliably be filled to $I_i = I_0$, since then the Chirikov resonance can sustain itself until all depot fuel is consumed even without dissipation. Dissipative control can be a major advantage, however, if filling the active particle's depot to its maximum capacity cannot be guaranteed. The question of how the energy depot can be filled leads us to our next Section.

\section{V. Collisions that can change the energy of $H_\text{D}$}

In Section III we allowed the momentum $p$ of the weight in Hamiltonian \eqref{eq:H_4dim} to change through stochastic collisions; the collisions were idealized as instantaneous momentum transfer between point-particles but our model for their effects was certainly consistent with Hamiltonian two-body mechanics. In this complementary section we examine a scenario in which random collisions instead change the canonical momentum $I$ of the depot subsystem \eqref{eq:H_4dim_realization_fuel}. In this section we will also further explore the relationship between Hamiltonian mechanics and the phenomenological concept of an active particle, by experimenting with merging the two kinds of description. Leaving a fully Hamiltonian description of energy harvesting for future work, we will now simply interrupt our Hamiltonian time evolution with certain randomly timed kicks, and then discuss the results and their meaning in our final Section.

\subsection{V.1 Motivating a model}
The kind of kicks which will now affect $I$ are motivated as follows. The basic idea is that since our active particle has an internal structure it should be able to exchange with its environment not only the kinetic energy of its active motion but also the internal energy of its depot. The depot energy will still be described by the $\braces{\alpha,I}$ system of \eqref{eq:H_4dim}-\eqref{eq:H_4dim_realization_coupling}, but we now consider this to be an effective description of a system which is really composed of multiple subsystems. Our final model will retain the continuous $I$ variable, but to develop the model we will initially imagine our depot in discrete terms, as an ensemble of some number $N$ of micro-depots. We seek a model for how our active particle might be able to gradually replenish its depot from the environment, by thinking of how it could refill one micro-depot at a time. 

In our discretized motivation for the model, therefore, we imagine an environment composed of small grains that can each hold enough energy to fill one empty micro-depot. For simplicity as we motivate the model, we allow each grain to have only two states: drained or full, denoted by $+$ or $-$. We propose that a fraction $P_+$ of grains are initially in the $+$ state while $(1-P_+)$ are in the $-$ state. In an effort to represent the time reversal symmetry that should be present in a more microscopic description, we will allow encounters with $+$ grains to fill an empty micro-depot, while encounters with $-$ grains will deplete the active particle's depot analogously, draining one micro-depot.

We further suppose that the chance of actually filling up a micro-depot during an encounter with a $+$ grain should be proportional to the number of empty micro-depots that are available to be filled, while likewise the chance of draining a micro-depot into a $-$ grain is proportional to the number of currently full micro-depots.

Having motivated our model with a discrete picture, then, we return to the continuous $I$ with the following stochastic process added to the Hamiltonian equations of motion for \eqref{eq:H_4dim}-\eqref{eq:H_4dim_realization_coupling}. Collisions occur at random times at the average rate $1/\bar{\tau}$, as in Section III. In each collision exactly three abrupt changes in $I$ are possible. With probability $P(I)$ the change is
 $I\to I+\delta I$, for a fixed constant $\delta I$ (as long as $I< I_0-\delta I$). With probability $\bar{P}(I)$ the change is instead $I\to I-\delta I$ (as long as $I>-I_0+\delta I$). With the remaining probability $1-P(I)-\bar{P}(I)$ there is then no change to $I$, because no empty or full micro-depot happened to connect with the respectively $+$ or $-$ grain. Since in the actual continuum description of the depot the proportions of full and empty micro-depots in our motivational picture should correspond to $(I_0\pm I)/(2I_0)$, the probabilities motivated by our micro-depot picture are therefore
\begin{align}
P(I) &= P_+ \frac{I_0-I}{2I_0},\nonumber\\
\bar{P}(I) &= (1-P_+)\frac{I_0 + I}{2I_0},\nonumber\\
1-P(I)-\bar{P}(I) &=\frac{1}{2} + \left(P_+-\frac{1}{2}\right)\frac{I}{2I_0}\;,
\end{align} 
where it may be noted that the chance of no change in $I$ satisfies $1/4\leq 1-P(I)-\bar{P}(I)\leq 3/4$. The rate of $+\delta I$ collisions in relation to $-\delta I$ collisions tends to increase as $I$ is lower, as if the active particle feeds more aggressively when it is hungry.

\subsection{V.2 Expectations from the $I$ kicks}
To anticipate what we may see for small but finite $\delta I$ and $\bar{\tau}$, it is straightforward to consider the Langevin-like differential equation which our random kicks would imply in the limit where $\delta I$ and $\bar{\tau}$ both go to zero while keeping their ratio fixed. In this limit no stochastic noise survives; we obtain a deterministic set of equations of motion with a damping-like term in the equation for $I$:
\begin{align} \label{eq:Idot}
\dot I = -\Gamma \braces{I - I_\text{s}} - \nach{H}{\alpha}\;,
\end{align}
where 
\begin{align}\label{eq:I_s}
\Gamma&=\frac{\delta I}{I_0\bar{\tau}}, & I_\text{s}=&\braces{2P_+ - 1}I_0.
\end{align}
The presence of $I_\text{s}$ makes this different from the otherwise similar case in Section IV. 

Since $I_\text{s} > -I_0$ for any $P_+>0$, the fuel depot without coupling $H_\text{C}$ to the active particle motion would always tend to reach equilibrium at some $I=I_\text{s}$ above its ground state $I=-I_0$. Whether $H_\text{D}$ increases or decreases then in general depends on the initial conditions of $I$. If $I$ is initially above $I_\text{s}$, it will decrease and increase otherwise. As opposed to the change in $p$ due to collisions, which always tends to lower the energy of $H_\text{M}$, and with that the total energy of \eqref{eq:H_4dim}, the change in $I$ can now lead to an energy flow into the system.

To anticipate what may happen when the fuel depot is put under load to drive the active phase of our Hamiltonian active particle, we can note that moving at the operating speed $v_\text{c}$ against the external force $f$ implies a power drain $v_\text{c} f = -\Omega \dot{I}$ from the depot. In the active phase we therefore approximate
\begin{align} \label{eq:Idota}
\dot I = -\Gamma \braces{I - I_\text{s}} - \frac{f}{k}\;,
\end{align}
which drives $I$ towards
\begin{equation}
I_{\text{a}} = I_\text{s} - \frac{f}{\Gamma k}\;.
\end{equation}
Depending on the various parameters it can happen that $I_\text{a}$ is too low---or too high!---for the force from $H_\text{C}$, which is proportional to $\sqrt{I_0^2-I^2}$, to overcome $f$ and keep the active particle moving. It can even be that $I_\text{a}<-I_0$, so that the depot is driven to $I=-I_0$ where the force from $H_\text{C}$ vanishes and the active phase always ends. 

It would seem from this analysis, however, that for sufficiently high $\Gamma$ and $P_+$ we will have $I_\text{a}$ well within the operating range of the active phase. The active particle should therefore be able to sustain its active phase indefinitely, by harvesting energy as it moves through its environment to continuously replenish its internal energy depot.

\subsection{V.3 Results with energy-harvesting $I$ kicks}
\begin{figure}[htb]
	\centering
	\includegraphics*[trim=239 311 276 356, width=0.45\textwidth]{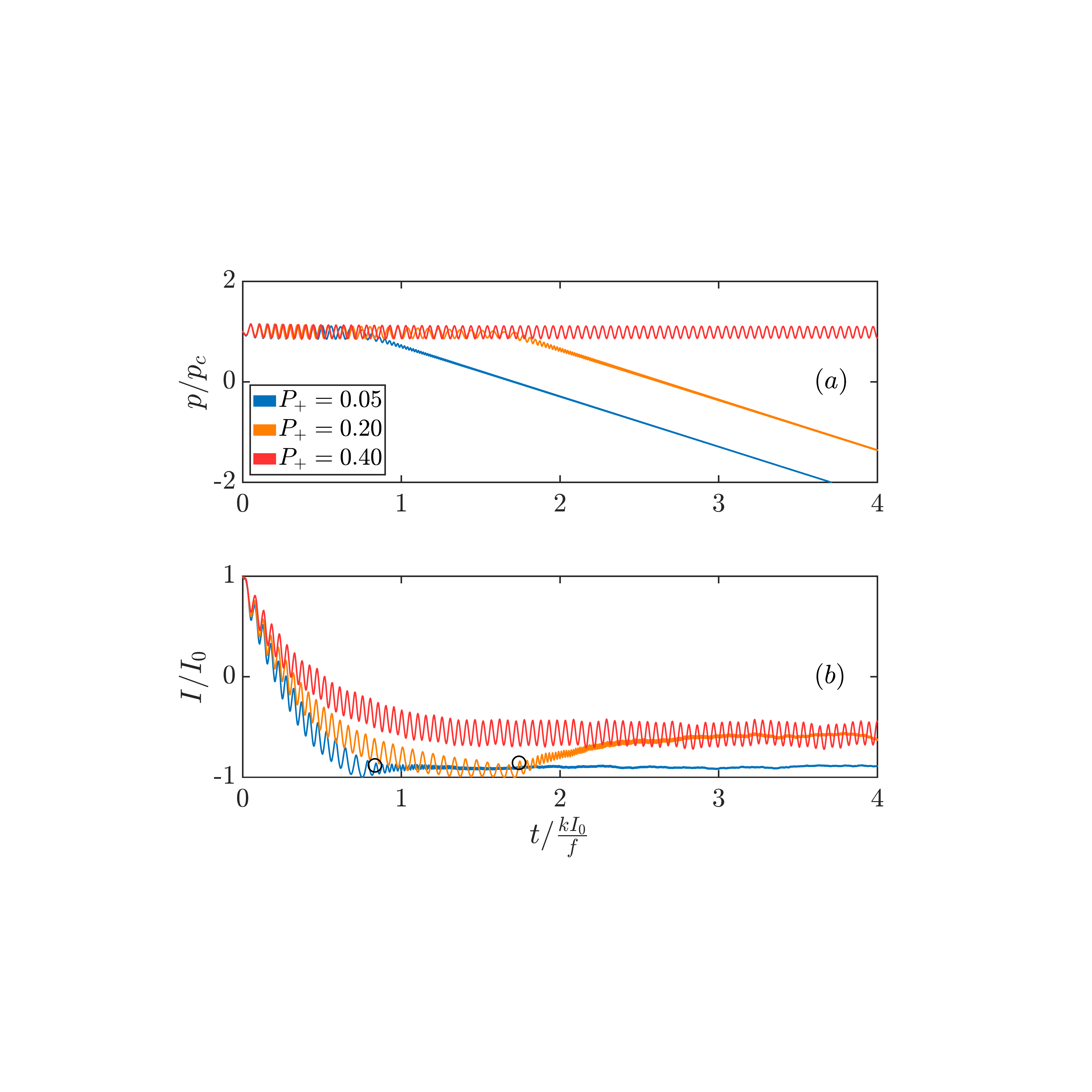}
	\caption{\label{fig:DemoThermalization} Momentum measures $p/p_\text{c}$ (a) and $I/I_0$ (b) for different probabilities $P_+$ of finding an environment particle in a '+' state plotted over the time measure $t/\frac{k I_0}{f}$. Collisions drain energy from the fuel system if $I$ is higher than the value $I_\text{s}$, and restore otherwise. If the restoring is high enough the weight might keep its critical momentum $p_\text{c}$ forever (red curve). Hence energy from the environment is used to lift the weight against a potential.}
\end{figure}

Such a perpetual active state can be seen in Fig.~\ref{fig:DemoThermalization}, which shows three representative trajectories for small but finite $\delta I$ and $\bar{\tau}$, differing only in the probability $P_+$, which is $0.05$ for the blue, $0.20$ for the orange, and $0.40$ for the red curves. The time measure $t/\frac{kI_0}{f}$ would be the same as the $-\lambda / I_0$ axis of our previous plots, if we now had no kicks.

Without the $I$ kicks the energy transfer would stop at $t/\frac{kI_0}{f} \approx 1$. From the momentum measure $p/p_c$ in Fig.~\ref{fig:DemoThermalization}(a), it can be seen that for the lowest simulated probability $P_+ = 0.05$ the energy transfer stops at $t/\frac{kI_0}{f} \approx 0.9$, \emph{before} it would have stopped without kicks. For $P_+ = 0.20$ the extended active phase makes the total amount of energy transferred to the weight increase by a factor of about $1.5$. And for $P_+ = 0.40$ the active phase goes on forever. 
It can be seen in Fig.~\ref{fig:DemoThermalization}(b) that in the eternal active phase $I$ never falls to the lower boundary $-I_0$, but in the cases where the energy harvesting only succeeds in extending the duration of the active phase, the active phase does end when $I$ falls too near the ground state $-I_0$. Since this ends the active phase, the power drain from sustaining the active phase is then lifted and $I$ settles back to its unloaded asymptotic value $I_\text{s}$, which is $-0.9I_0$ in the case $P_+ = 0.05$ and $-0.6 I_0$ for $P_+ = 0.2$.

Real active particles which are more complicated than this simple model, including humans, can indeed harvest energy from their environments in order to sustain active motion for much longer than their internal depots would allow in isolation. Sufficient consumption efficiency is required, in an environment that is sufficiently rich in available energy. Our partially phenomenological model is thus at least qualitatively realistic. We now conclude by discussing what we have learned from how our models have worked.

\section{VI. Discussion}

Whatever an active particle may actually have for an engine, it must be a Hamiltonian system of some kind, with dynamical phases at least qualitatively analogous to those in our model. In such a dynamical system, especially if it is microscopic, all of the subsystems may well be affected by damping and noise that act on any of them. So what can happen to dynamical systems of this unusual but important kind when they are immersed in a dissipative environment?

\subsection{VI.1 Robust mechanisms}
In Section II of this paper we reviewed a well-defined microscopic model for a Hamiltonian active particle, previously presented as an isolated Hamiltonian system from \cite{ClassicalDaemon}. In Section III we confirmed the important result that even though this active particle model relies on the rather exotic nonlinear mechanism of a Chirikov resonance, it is robust enough to continue functioning in a dissipative and noisy environment. Too much noise can disrupt it, and too much dissipation can reduce its efficiency or even stall it completely, but its tolerances for dissipation and noise are non-zero. The microscopic mechanisms of real active particles are undoubtedly more complicated but for the project of nonetheless understanding them from physical first principles, this result is encouraging.

\subsection{VI.2 Dissipative control}
In Section IV we then removed noise to focus only on environmental dissipation, and found that dissipation can actually \emph{enhance} the efficiency with which a Hamiltonian active particle exploits its internal depot, by enhancing its own automatic control of the dynamical process that sustains active motion.

This is more than just the environment acting as a heat sink in the Carnot sense. The analogs of cold and hot Carnot reservoirs in the Hamiltonian daemon are discussed in \cite{ClassicalDaemon}, where the single rotor degree of freedom $(\alpha,I)$ of our depot is realized as the Schwinger representation of angular momentum with two harmonic oscillators of different frequency. The role of dissipation in enhancing the efficiency of the daemon is more comparable to the role of cooling a macroscopic engine in preventing excess micromotion (typically leading to thermal expansion that increases friction and ultimately makes parts jam together) in the engine's own moving parts. A Carnot cycle is an idealized picture of a heat engine in which the engine's moving parts are assumed not to deviate from their intended motion in any way, and not to absorb any heat. In real engines the moving parts, which are supposed to enforce the thermodynamic cycle mechanically, can in fact overheat and fail to perform as designed. Approaching closer to the ideal limit of a Carnot engine requires cooling the real engine's operating parts, as well as merely extracting heat from the working fluid into the cold reservoir within the ideal scheme.

In the daemon the micromotion of the system components, shown in our phase space plots, is itself the cyclic motion of the active phase engine. It is nonetheless true that what ultimately limits the efficiency of the Hamiltonian daemon \cite{ClassicalDaemon} is having too much micromotion. The difference from the overheating problem in macroscopic engines is that in the isolated daemon micromotion does not increase, but is simply conserved, under the adiabatic theorem. The efficiency of the daemon can be increased by damping this undesirable micromotion away. If we consider that dissipation can enhance engine efficiency by suppressing micromotion in internal components that would limit efficiency, though, then the role of dissipative control in enhancing the efficiency of a Hamiltonian active particle can indeed be considered a microscopic analog of cooling an engine. 

Understanding this potential benefit of the dissipative environment for active Brownian particles has been a major result of our paper:
\begin{center}
\emph{Dissipation can increase the total amount of energy transfer from the internal depot to the motional degree of freedom.}
\end{center}

\subsection{VI.3 Energy harvesting}
In Section V we showed that phenomenological and Hamiltonian approaches to active particle modelling can be compatible, at least in yielding realistic results. In particular we showed that it is possible to model the continuous replenishment of the energy depot of a Hamiltonian active particle, even to the limit where energy harvesting can sustain active motion indefinitely.

Careful consideration of our procedures in Section V also shows some limitations of the phenomenological approach, however. Any Hamiltonian system is intrinsically possible under our most basic understanding of the laws of nature, but phenomenologically guessed evolution rules may potentially violate those basic laws, even when this is not obvious. Within the seemingly reasonable assumptions of our motivational models of micro-depots and grains, for example, it is not clear that our active particle's perpetual motion could not be sustained by harvesting energy from an environment in thermal equilibrium at a positive temperature. This would be a perpetual motion machine of the Second Kind, forbidden by the Second Law of Thermodynamics.

The perpetual active motion that we showed in Section V does not have to violate the Second Law, if it presumes an environment which is far from thermodynamical equilibrium or has a negative temperature. The problem with the phenomenological model is not that it is necessarily qualitatively wrong, but only that it does not reliably indicate when it may be qualitatively wrong. Identifying such problems can require the kind of careful reasoning that is typically needed to resolve the apparent paradoxes of Maxwell's Demon scenarios, of which this energy harvesting model is potentially one.

Instead of inductively searching for clues to the mistaken assumptions in that way, the alternative procedure that we mean to suggest with the whole of this paper is to formulate the entire problem within Hamiltonian mechanics, as at the start of this paper, and proceed deductively to work out what works, or does not, when consistency with fundamental mechanics is built-in from the start.

For many purposes that may not be needed. When enough is known about how systems in fact behave, phenomenological models can be checked against that knowledge. To understand fully how active particles work, and know whether previous limits can in fact be exceeded, may require a deeper understanding, however. A fully Hamiltonian description of energy harvesting by an active particle is a project in progress.

The authors acknowledge support from State Research Center OPTIMAS and the Deutsche Forschungsgemein- schaft (DFG) through SFB/TR185 (OSCAR), Project No. 277625399.

\section{Appendix:}
\subsection{A. Multiple time scale analysis of the dissipative daemon}
Here we do not begin with the effective 2D phase space of Section II, but consider the full canonical Hamiltonian equations of motions derived from \eqref{eq:H_4dim} with the addition of a standard viscous damping term on the particle momentum $p$, with damping coefficient $\gamma$:
\begin{align}
\dot{p} &= -f-\varepsilon k\Omega\sqrt{I_0^2-I^2}\sin(kq-\alpha)-\gamma p\;,\nonumber\\
\dot{q} &= \frac{p}{M}\;,\nonumber\\
\dot{I} &= \varepsilon\Omega\sqrt{I_0^2-I^2}\sin(kq-\alpha)\;,\nonumber\\
\dot{\alpha} &= \Omega + \varepsilon\frac{\Omega I}{\sqrt{I_0^2-I^2}}\cos(kq-\alpha)\;.
\end{align}
In the active phase of evolution we will have $p\sim M\Omega/k$ as explained above, and so active motion will only be possible if the total retarding force $f+\gamma M\Omega/k$ is less than the maximum force from $H_\text{C}$ $\varepsilon k\Omega I_0$.

We now introduce dimensionless variables involving explicit dimensionless factors $\varepsilon\ll1$ and $\Delta = \sqrt{M\Omega/(I_0 k^2)}\leq 1$:
\begin{align}
t &=\frac{\Delta}{\sqrt{\varepsilon}}\frac{\tilde{t}}{\Omega}\;, & q &= \frac{Q+\alpha}{k}\;, \nonumber\\
f &=\varepsilon I_0 k \Omega F\;,   & p&= \frac{M\Omega}{k}\Big(1+\frac{\sqrt{\varepsilon}}{\Delta}P\Big)\;,\nonumber\\
\gamma &= \frac{\varepsilon I_0 k^2}{M}\Gamma\;, & I  &= X I_0
\;.
%p &= \frac{M\Omega}{k}\Big(1+\frac{\sqrt{\varepsilon}}{\Delta}P\Big)\nonumber\\
%q &= \frac{Q+\alpha}{k}\nonumber\\
%I & = X I_0\;.
\end{align}
With these definitions we expect the active phase to be possible for $F+\Gamma < 1$.
This yields the dimensionless equations of motion
\begin{align}\label{dimEQ}
\frac{dQ}{d\tilde{t}}&= P - \sqrt{\varepsilon}\Delta \frac{X \cos Q}{\sqrt{1-X^2}}\;,\nonumber\\
\frac{dP}{d\tilde{t}} &= - (F+\Gamma) - \sqrt{1-X^2}\sin Q -  \frac{\sqrt{\varepsilon}}{\Delta}\Gamma P\;,\nonumber\\
\frac{dX}{d\tilde{t}} &= \sqrt{\varepsilon}\Delta\sqrt{1-X^2}\sin Q\;.
\end{align}
We do not need the additional equations of motion for $q$ or $\alpha$ separately, given $Q$, since these are trivial; the closed system of equations \eqref{dimEQ} determines the full evolution.

We then exploit the assumed smallness of $\varepsilon$ by implementing the standard multiple time scale perturbation theory. We interpret the dimensionless time coordinate $\tilde{t}$ as a line of small slope in the plane of fast and slow times $\tau$ and $T$:
\begin{align}
(\tau,T) &= (1,\sqrt{\varepsilon})\tilde{t}\;,\nonumber\\
Q(\tau,T) &= \sum_{n=0}^\infty Q_n(\tau,T)\varepsilon^{n/2}\;,\nonumber\\
P(\tau,T) &= \sum_{n=0}^\infty P_n(\tau,T)\varepsilon^{n/2}\;,\nonumber\\
X(\tau,T) &= \sum_{n=0}^\infty X_n(\tau,T)\varepsilon^{n/2}\;.
\end{align}
The evolution of the dynamical variables is only physically meaningful along the time line $(\tau,T) = (1,\sqrt{\varepsilon})\tilde{t}$; we are free to invent arbitrary evolution in the rest of the $(\tau,T)$ plane, and as usual in multiple time scale analysis we use this freedom to force all non-periodic evolution to occur in $T$ rather than $\tau$. Otherwise we simply expand order by order in $\sqrt{\varepsilon}$ as in ordinary classical Hamiltonian perturbation theory.

At order $\varepsilon^0$ we find 
\begin{align}\label{QP0}
\frac{\partial Q_0}{\partial \tau} &= P_0\;,\nonumber\\
\frac{\partial P_0}{\partial \tau}&= -(F+\Gamma) - \sqrt{1-X_0^2}\sin Q_0
\end{align}
plus the conclusion that $X_0=X_0(T)$ cannot depend on the fast time $\tau$. These equations of motion have the first integral
\begin{equation}
E_0(T) = \frac{P_0^2}{2} + (F+\Gamma)Q_0-\sqrt{1-X_0(T)^2}\cos Q_0\;,
\end{equation}
recognizable of course as an energy which is independent of $\tau$ but depends on $T$ in a manner as yet undetermined. 

Identifying this first integral $E_0(T)$ may be considered as solving the dependence of $Q_0$ and $P_0$ on $\tau$. As $\tau$ increases while $T$ is held fixed, $Q_0(\tau,T)$ and $P_0(\tau,T)$ trace out orbits in their 2D phase space. In particular for any value of $X_0(T)=X$ we can identify the zeroth-order unstable fixed point $(Q_0,P_0) = (Q_S(X),0)$ for
\begin{equation}
Q_S(X)=-\sin^{-1}\Big(\frac{F+\Gamma}{\sqrt{1-X^2}}\Big)\;.
\end{equation}
The energy $E_0$ at this fixed point $(Q_0,P_0) \to (Q_S(X),0)$ is the energy of the separatrix $E_S(X)$ ; the single closed orbit of $Q_0,P_0$ in dependence on $\tau$ with fixed $T$ and $E_0(T) = E_S(X(T))$ is the separatrix itself at slow time $T$. All orbits with
\begin{equation}
E_0 < E_S(X) = (F+\Gamma)Q_S(X) + \sqrt{\frac{1-X^2 - (F+\Gamma)^2}{1-X^2}}
\end{equation}
evolve in $\tau$ as closed orbits in the $Q_0,P_0$ plane, inside the separatrix. The separatrix as well as the individual orbits all depend on $T$ (through $X\to X_0(T)$) as well as on $\tau$; with the $\tau$-dependence solved we now turn to finding this $T$-dependence---a question analogous to determine climate change, having solved for the weather.

\begin{widetext}To determine the $T$-dependence of all our variables it will turn out to be useful to define the area enclosed in the $(Q_0,P_0)$ phase space by a closed $\tau$-orbit with energy $E_0 = E\leq E_S(X)$, namely the \emph{action}
\begin{align}
S_0(E,X) := \oint\!d\tau P_0^2 = \oint dQ P_0(Q,E,X)
 = 2\sqrt{2}\int_{Q_-(E,T)}^{Q_+(E,T)}\!\!\!\!\!\!dQ\,\sqrt{E - (F+\Gamma)Q+\sqrt{1-X^2}\cos Q}\;,
\end{align}
where $\oint\!d\tau$ denotes the integral over a full period in $\tau$, and $Q_\pm(E,T)$ are the two values of $Q$ (modulo $2\pi$) at which the integrand vanishes (the turning points where $\partial_\tau Q_0 = P_0$ vanishes for $E_0 = E$). The reason we care about this action is that it will turn out below to have a simple evolution in $T$. The evolution in $T$ of $S_0(E,T)$ determines implicitly the $T$-dependence of $E_0$, and this in turn fixes the $T$-dependence of $Q_0$ and $P_0$.

Anticipating its use below, therefore, we compute the derivative of $S_0$ with respect to $T$ and find
\begin{align}\label{dTS}
\partial_T S_0\big(E_0(T),X_0(T)\big) &= 2 \int_{Q_-(E_0,T)}^{Q_+(E_0,T)}\!dQ\,\frac{\partial_T E_0(T) - \big(1-X_0(T)^2\big)^{-1/2}\,X_0\partial_T X_0\,\cos Q}{\sqrt{2\Big(E_0 - (F+\Gamma)Q+\sqrt{1-X_0(T)^2}\cos Q\Big)}}\nonumber\\
&\equiv \oint\!d\tau\,\Big(\partial_T E_0(T) + \cos\big(Q_0(\tau,T)\big)\partial_T \sqrt{1-X_0(T)^2}\Big)\nonumber\\
&\equiv \oint\!d\tau\,\Big(P_0\partial_T P_0 + [F+\Gamma + \sqrt{1-X_0^2}\sin Q_0]\partial_T Q_0\Big)\;,
\end{align}
where in the first line there is no contribution from the derivatives of the integration limits because the integrand vanishes at them, and the $\oint$ symbol in the second and third lines denotes integration over a single period of the evolution in $\tau$.

To obtain the actual $T$-dependence $S_0$ according to the equations of motion, we proceed to order $\varepsilon^{1/2}$ in our perturbative theory. We first examine
\begin{align}\label{X1}
\partial_\tau X_1 &= -\partial_T X_0 + \Delta\sqrt{1-X_0^2}\sin Q_0 =-X'_0(T) - \Delta (\partial_\tau P_0 + F +\Gamma) \;.
\end{align}
The constraint that all steadily growing or decreasing behavior must appear as dependence on $T$ rather than on $\tau$ means that our functions must all depend on $\tau$ periodically (though the period can depend on energy arbitrarily), and so whatever this periodic evolution in $\tau$ may be, we can integrate both sides of \eqref{X1} with respect to $\tau$ over one full period to obtain
\begin{equation}\label{X00}
X'_0(T) = - \Delta (F +\Gamma)\;.
\end{equation}
Expressed in the dimensionless variables this is the result described in our main text, that the fuel depot $I$ steadily decreases, when the particle is active, to provide the power needed to sustain steady motion against the external and dissipative force.

Inserting \eqref{X00} in \eqref{X1} then yields
\begin{equation}
X_1(\tau,T) = -\Delta P_0(\tau,T) + x_1(T)
\end{equation}
for some function $x_1(T)$.

We further find
\begin{align}
\frac{\partial Q_1}{\partial \tau}-P_1&=-\partial_T Q_0 - \Delta \frac{X_0 \cos Q_0}{\sqrt{1-X_0^2}}\;,\label{Q1}\\
\frac{\partial P_1}{\partial \tau} + \sqrt{1-X_0^2}\cos(Q_0) Q_1 &= - \partial_T P_0 - \frac{\Gamma}{\Delta}P_0 + \frac{X_0 X_1 \sin(Q_0)}{\sqrt{1-X_0^2}} \nonumber\\
&= - \partial_T P_0 - \frac{\Gamma}{\Delta}P_0 - \frac{X_0(T)[x_1(T) - \Delta P_0]}{1-X_0(T)^2}\left(\partial_\tau P_0 + F +\Gamma\right)\label{P1}\;.
\end{align}

We then note that integration by parts and Eqns.~\eqref{QP0} imply
\begin{equation}\label{iden}
\oint\!d\tau\,\left[\left(\sqrt{1-X_0^2}\sin (Q_0) + F+\Gamma\right) [\partial_\tau Q_1 - P_1] + P_0[\partial_\tau P_1 + \sqrt{1-X_0^2}\cos(Q_0) Q_1]\right]
 = 0
 \end{equation}
 when the integration is over any full period in $\tau$.  We insert \eqref{Q1} and \eqref{P1} in \eqref{iden} and discard total derivatives with respect to $\tau$ from the full-period integral, then compare the result with \eqref{dTS} to reveal the surprisingly simple conclusion
 \begin{equation}\label{S0T}
 \frac{dS_0}{dT} = - \frac{\Gamma}{\Delta}S_0(T)\;.
  \end{equation}
This may be recognized as the generalization to include dissipation of the classical adiabatic theorem, according to which (with $\Gamma =0$) the action is a so-called \emph{adiabatic invariant}, meaning that even though the energy and all other dynamical variables may have slow trends that gradually accumulate to significant change, these trends will conspire to keep their particular combination in the action constant even over long times. 
 \end{widetext}
 
 With dissipation instead the action decreases exponentially. Restoring physical time units reveals that up to corrections of order $\sqrt{\varepsilon}$ (the higher order perturbative terms $n>0$), which remain that small even over long times, the action defined by $S_0$ decreases exponentially at the rate $\gamma$, \emph{i.e.}
\begin{equation}
S_0(t) = S_0(0) e^{-\gamma t}\;.
\end{equation}
 
To determine the efficiency with which the active particle uses its depot energy before the active phase ceases, we use the fact that the active phase can only begin with the system entering an orbit just inside the separatrix when $X$ has the initial value $I_i/I_0$. (We do not consider initial conditions that have been fine-tuned, through some external process not described by our Hamiltonian, to start the system in its active phase. So if the system is deeper inside the separatrix, then it must already have been inside the separatrix for some time, meaning that the beginning of the active phase must have been earlier. The active phase therefore always begins with the system on the separatrix as it exists for $X(T=0) = I_i/I_0$.) 

While the system is in the active phase, the separatrix grows or shrinks as $X(T) = X(0)-\Delta(\Gamma+F)T$ changes, while the system's orbit shrinks under dissipation. The active phase ends when the area enclosed by the system's orbit exceeds the area inside the separatrix. The separatrix area for a given value of $X(T)$ is $S_{sep}(X) = S_0(E_S(X),X)$, and by identifying the scaled duration of the active phase as $T_f = [X_0(T_f)-X_0(0)]/[(\Gamma + F)\Delta]$ we find the remaining fuel level $I_f = I_0 X_0(T_f)$ at the end of the active phase from the condition
\begin{equation}\label{condit}
S_\text{sep}(I_i/I_0) e^{-\frac{\Gamma}{\Gamma+F}\frac{I_i}{I_0\Delta^2}} = S_\text{sep}(I_f/I_0)e^{-\frac{\Gamma}{\Gamma+F}\frac{I_f}{I_0\Delta^2}}\;.
\end{equation}

To see that \eqref{condit} is a well-defined condition which can be solved numerically to determine $X_f(X_i)$, we can note that in the limit of small $\Gamma + F$ we can approximate $Q_S(X)=Q_-(X)\doteq -\pi\doteq -Q_+(X)$ and $E_S(X)\doteq \sqrt{1-X^2}$ to obtain the explicit analytical formula
\begin{align}
S_\text{sep}(X) &\doteq 2\sqrt{2}\braces{1-X^2}^{1/4}\int_{-\pi}^{\pi}\!\!\!\!\!\!dQ\,\sqrt{1 + \cos Q}\nonumber\\
&=16\braces{1-X^2}^{1/4}\;.
\end{align}
For general $F$ and $\Gamma$ the numerical task of solving \eqref{condit} is a bit harder because $S_\text{sep}(X)$ itself requires numerical integration. This is how we obtained the adiabatically predicted efficiency curves in our main text.

Regardless of exactly how $S_\text{sep}(X)$ behaves it is easy to note that $S_\text{sep}(X) =S_\text{sep}(-X)$ and that $S_\text{sep}(\pm 1) = 0$, so that $I_i = 1$ implies $X_f=-1$ and all fuel is used regardless of $\Gamma/F$, while for $\Gamma=0$ and $I_i<I_0$ we have $I_f = -I_i$ (and the active phase can never begin for $I_i < 0$ \cite{ClassicalDaemon}). For $I_i < 1$ the active particle without dissipation is inefficient, in that it is unable to use all its potentially available fuel. Whenever dissipation is strong enough that
\begin{equation}
\exp\left[-\frac{\Gamma}{\Gamma +F}\frac{1}{\Delta^2}\right]\doteq 0\;,
\end{equation}
however, we will have $X_f \doteq -1$ regardless of $X_i$, meaning that all fuel is used. When $\Delta^2$ is small even a very slight amount of dissipation can dramatically boost the efficiency with which the Hamiltonian active particle can exploit its energy depot.


\begin{thebibliography}{10}

%\bibitem{Prestin} Zheng, J., Shen, W., He, D. et al., Prestin is the motor protein of cochlear outer hair cells, \textit{Nature} \textbf{405}, 149-155 (2000)

\bibitem{Broken} F.S. Gnesotto, F. Mura, J. Gladrow, and C.P. Broedersz, Broken detailed balance and non-equilibrium dynamics in living systems: a review, \textit{Rep. Prog. Phys.} \textbf{81} 066601 (2018)

\bibitem{Noisy} P. Reimann, Brownian motors: noisy transport far from equilibrium, \textit{Phys. Rep.} \textbf{361} 57–265 (2002)

\bibitem{Ratchet} B. Lau, O. Kedem, J. Schwabacher, D. Kwasnieskiab and  E. A. Weiss, An introduction to ratchets in chemistry and biology, \textit{Mater. Horiz.} \textbf{4}, 310-318 (2017)

\bibitem{BrownMotor} P. Hänggi, F. Marchesoni, and F. Nori, Brownian motors, \textit{Ann. Phys.} \textbf{14}(1-3) 51-70 (2005)

\bibitem{Extone} K. Villa and M. Pumera, Fuel-free light-driven micro/nanomachines: artificial active matter mimicking nature \textit{Chem. Soc. Rev.} \textbf{48}, 4966, (2019)

\bibitem{Exttwo} T. Xu, L. Xu, and X. Zhang, Ultrasound propulsion of micro-/nanomotors, \textit{Appl. Mater. Today} \textbf{9}, 493–503 (2017) 

\bibitem{Design} P. Pietzonka, É. Fodor, C. Lohrmann, M. E. Cates, and U. Seifert, Autonomous Engines Driven by Active Matter: Energetics and Design Principles \textit{Phys. Rev. X} \textbf{9}, 041032 (2019)

\bibitem{ActStat} S. Ramaswamy, The Mechanics and Statistics of Active Matter, \textit{Annu. Rev. Condensed Matter Phys.} \textbf{1}, 323-345 (2010)

\bibitem{Complex} F. Schweitzer, W. Ebeling, and B. Tilch, Complex Motion of Brownian Particles with Energy Depots, \textit{Phys. Rev. Lett.} \textbf{80}, 5044 (1998) 

\bibitem{Uphill} F. Schweitzer, B. Tilch, and W. Ebeling, Uphill motion of active brownian particles in piecewise linear potentials, \textit{Eur. Phys. J. B} \textbf{14}, 157–168 (2000)

\bibitem{ActiveMotion} Y. Zhang, C. Kim, and K.-J.-B. Lee, Active motions of Brownian particles in a generalized energy-depot model, \textit{New J. Phys.} \textbf{10}, 103018 (2008)

\bibitem{Canon} A. Gl\"uck, H. H\"uffel, and S. Ilijić, Canonical active Brownian motion, \textit{Phys. Rev. E} \textbf{79} 021120 (2009)

\bibitem{DirMot} B. Tilch, F. Schweitzer, and W. Ebeling, Directed motion of Brownian particles with internal energy depot, \textit{Physica A} \textbf{273} 294-314 (1999)

\bibitem{ChemFree} T. Dittrich and N.A. Naranjo, Directed transport in a ratchet with internal and chemical freedoms, \textit{Chemical Physics} \textbf{375} 486–491 (2010)

\bibitem{ActiveReview} P. Romanczuk, M. Bär, W. Ebeling, B. Lindner, and L.Schimansky-Geier, Active Brownian Particles. From Individual to Collective Stochastic Dynamics, \textit{Eur. Phys. J. Spec. Top.} \textbf{202}, 1-162 (2012)

\bibitem{Assym} P.J. Park and K.-J.-B. Lee, A modified active Brownian dynamics model using asymmetric energy conversion and its application to the molecular motor system, \textit{J. Biol. Phys.} \textbf{39}, 439–452 (2013)

\bibitem{selfdriven} T. Vicsek, A. Czirók, E. Ben-Jacob, I. Cohen, and O. Shochet, Novel Type of Phase Transition in a System of Self-Driven Particles, \textit{Phys. Rev. Lett.} \textbf{75}, 1226 (1995).

\bibitem{ActiveSus} K. Kanazawa, T. Sano, A. Cairoli, and A. Baule, Loopy Lévy flights enhance tracer diffusion in active suspensions, \textit{Nature} \textbf{579}, 364–367 (2020). 

\bibitem{Information} U. Khadka, V. Holubec, H. Yang, and F. Cichos, Active particles bound by information flows. \textit{Nat. Comm.} \textbf{9}, 3864 (2018)

\bibitem{Coop} C. Touya, T. Schwalger, and B. Lindner, Relation between cooperative molecular motors and active Brownian particles, \textit{Phys. Rev. E} \textbf{83}, 051913 (2011)

\bibitem{Active_ComplexEnvironment} C. Bechinger, R. Leonardo, H. Löwen, C. Reichhardt, G. Volpe, and G. Volpe, Active particles in complex and crowded environments, \textit{Rev. Mod. Phys.} \textbf{88}, 045006 (2016)

\bibitem{Cognition} Patrick McGivern, Active materials: minimal models of cognition?, \textit{Adaptive Behavior} (2019)

\bibitem{Limits} R. Marsland III and J. England, Limits of predictions in thermodynamic systems: a review, \textit{Rep. Prog. Phys.} \textbf{81} 016601 (2018)

\bibitem{Seifert} U. Seifert, Stochastic thermodynamics, fluctuation theorems and molecular machines, \textit{Rep. Prog. Phys.} \textbf{75} 126001 (2012)

\bibitem{ActStoch} C. Ganguly and D. Chaudhuri, Stochastic thermodynamics of active Brownian particles, \textit{Phys. Rev. E} \textbf{88}, 032102 (2013)

\bibitem{ActiveEntropy} P. Pietzonka and U. Seifert, Entropy production of active particles and for particles in active baths, \textit{J. Phys. A: Math. Theor.} \textbf{51} 01LT01 (2018)

\bibitem{Stochastic_Energetics} K. Sekimoto, Stochastic energetics, \textit{Lect. Notes Phys.} \textbf{799} (2010)

\bibitem{Need} M.E. Cates, Diffusive transport without detailed balance in motile bacteria: does microbiology need statistical physics?,\textit{Rep. Prog. Phys.} \textbf{75} 042601 (2012)

\bibitem{Challenging} A.-S. Smith, Physics challenged by cells, \textit{Nat. Phys.}, \textbf{6}, 726-729 (2010)

\bibitem{Opinion} N.L. Abbott and O.D. Velev, Active particles propelled into researchers’ focus, \textit{Curr. Opin. Colloid Interface Sci.} \textbf{21}, 1–3 (2016).

\bibitem{ClassicalDaemon} L. Gilz, E. Thesing, and J.R. Anglin, Hamiltonian analogs of combustion engines: A systematic exception to adiabatic decoupling, \textit{Phys. Rev. E} \textbf{94}, 042127 (2016) 

%\bibitem{QuantumDaemon} L. Gilz, E. Thesing, and J.R. Anglin, Quantum Hamiltonian daemons: Unitary analogs of combustion engines, \textit{Phys. Rev. E} \textbf{96}, 012119 (2017) 

\bibitem{Chirikov} B.V. Chirikov, Research concerning the theory of non-linear resonance and stochasticity,
\textit{Print N 267 Institute of Nuclear Physics, Novosibirsk}, (1969)

\bibitem{Liouvillepaper} T. Eichmann, E.P. Thesing, and J.R. Anglin, Engineering separatrix volume as a control technique for dynamical transitions, \textit{Phys. Rev. E} \textbf{98}, 052216 (2018)

\bibitem{Piecewise_deterministic} M. Davis and M. Dempster, Piecewise-deterministic Markov processes: A general class of non-diffusion stochastic models, \textit{J. of the Royal Stat. Soc. Series B: Methodological}, \textbf{46}, 353-388 (1984)

\bibitem{Goldstein} H.P. Goldstein, P. Charles, and J.L. Safko, \textit{Classical mechanics}, Addison Wesley (2002)



%\bibitem{Ralf} R. B\"urkle, A. Vardi, D. Cohen, and J.R. Anglin, Probabilistic Hysteresis in Integrable and Chaotic Isolated Hamiltonian Systems, \textit{Phys. Rev. Lett.} \textbf{123}, 113101 (2019)

%\bibitem{ExploitNoise} Piero Olla, Pros and cons of swimming in a noisy environment, \textit{Phys. Rev. E} \textbf{89}, 032136 (2014)

%\bibitem{Cappi} R.~Cappi and M.~Giovannozzi, Novel Method for Multiturn Extraction: Trapping Charged Particles in Islands of Phase Space, \textit{Phys. Rev. Lett.} \textbf{88}, 104801 (2002)

%\bibitem{Neishtadt} A. I.~Neishtadt, Passage through a separatrix in a resonance problem with a slowly-varying parameter, \textit{J. Appl. Math. Mech.} \textbf{39}, 594-605 (1975) 

%\bibitem{Autonomous_motor} H.~C.~Fogedby and A.~Imparato, A minimal model of an autonomous thermal motor, \textit{Europhys. Lett.} \textbf{119}, 50007 (2017)



\end{thebibliography}
\end{document}